\documentclass[aps, pre, reprint, unsortedaddress, superscriptaddress]{revtex4-1}
\usepackage{graphicx}
\usepackage{hyperref}
\usepackage{amsmath}
\usepackage{amssymb}
\usepackage{bm}

\begin{document}
\title{Momentum transport and nonlocality in heat-flux-driven magnetic reconnection in high energy density plasmas}
\author{Chang Liu}
\affiliation{Department of Astrophysical Sciences, Princeton University, Princeton, New Jersey 08544, USA}
\affiliation{Princeton Plasma Physics Laboratory, Princeton, New Jersey 08540, USA}
\author{William Fox}
\affiliation{Princeton Plasma Physics Laboratory, Princeton, New Jersey 08540, USA}
\author{Amitava Bhattacharjee}
\affiliation{Department of Astrophysical Sciences, Princeton University, Princeton, New Jersey 08544, USA}
\affiliation{Princeton Plasma Physics Laboratory, Princeton, New Jersey 08540, USA}
\author{Alexander G. R. Thomas}
\affiliation{Department of Physics, Lancaster University, Lancaster, United Kingdom LA1 4YB}
\affiliation{Department of Nuclear Engineering and Radiological Sciences, University of Michigan, Ann Arbor, Michigan 48109, USA}
\author{Archis Joglekar}
\affiliation{Department of Physics and Astronomy, University of California, Los Angeles, California 90095, USA}
\affiliation{Department of Nuclear Engineering and Radiological Sciences, University of Michigan, Ann Arbor, Michigan 48109, USA}

\date{\today}

\begin{abstract}

Recent theory has demonstrated a novel physics regime for magnetic reconnection in high-energy-density plasmas where the magnetic field is advected by heat flux via the Nernst effect. Here we elucidate the physics of the electron dissipation layer in this regime. Through fully kinetic simulation and a generalized Ohm's law derived from first principles, we show that momentum transport due to a nonlocal effect, the heat-flux-viscosity, provides the dissipation mechanism for magnetic reconnection. Scaling analysis and simulations show that the reconnection process is comprised of a magnetic field compression stage and quasi-steady reconnection stage, and the characteristic width of the current sheet in this regime is several electron mean-free-paths. These results show the important interplay between nonlocal transport effects and generation of anisotropic components to the distribution function.

\end{abstract}

\maketitle

\section{Introduction}

Magnetic fields in high-energy-density (HED) plasmas are of interest as the field can modify 
and direct electron heat flux and therefore determine the energy confinement properties of these plasmas.
Strong, Mega-Gauss-scale (MG-scale) magnetic fields can be generated in laser-target interactions by 
a number of mechanisms including the Biermann battery effect 
\cite{stamper_spontaneous_1971}, plasma instabilities in coronal plasmas\cite{rygg_proton_2008},
the Rayleigh-Taylor instability \cite{gao_magnetic_2012}, and the Weibel instability \cite{fox_filamentation_2013,huntington_observation_2015}.
The presence of magnetic fields improves energy confinement in hohlraums \cite{glenzer_thomson_1999}.
Magnetic fields can also be applied externally with pulsed-power systems,
underlying the MagLIF fusion concept \cite{slutz_pulsed-power-driven_2010}, 
and, for example, have been shown to improve direct-drive
fusion performance on OMEGA \cite{chang_fusion_2011}. 

A novel effect in many HED plasmas is that
the magnetic field can itself be advected by the heat 
flux, via the so-called
Nernst effect in the generalized Ohm's law (GOL) \cite{haines_heat_1986,kho_nonlinear_1985,ridgers_magnetic_2008}.  
The Nernst effect arises from the $v^{-3}$ velocity dependence of the collision frequency in plasmas;
intuitively, the magnetic field appears frozen to the low-collisionality, hot
population of electrons, but diffuses across the compensating colder return current, with the 
net effect that the field advects parallel to the heat flux.
Several experimental results have demonstrated the importance of the 
Nernst effect in HED regimes \cite{froula_quenching_2007,willingale_fast_2010,lancia_topology_2014,gao_precision_2015}, with promising agreement obtained between experiment and simulation.  
Nevertheless, correctly simulating the evolution of the magnetic field in these systems remains a challenge due to the coupling of the magnetic field to the heat flux, which can be nonlocal in character\cite{bell_elecron_1981,kho_nonlinear_1985,luciani_magnetic_1985}.

The evolution of the magnetic field in these HED systems can be further determined by magnetic reconnection driven by collision of opposing magnetic fields\cite{nilson_magnetic_2006,li_observation_2007,zhong_modelling_2010}.  Magnetic reconnection then affects the self-organization of plasma profiles by modifying the plasma transport processes, even in high-beta regimes. 
%Magnetic fields from multiple expanding plasmas can be advected and finally collide, leading to the interaction of 
%opposing magnetic fields and subsequent magnetic reconnection 
%\cite{nilson_magnetic_2006,li_observation_2007,zhong_modelling_2010}. Magnetic reconnection can be important for the evolution of magnetic field, changing the heat flux and other transport, and affecting plasma self-organization.
%%Magnetic reconnection plays an important role in laboratory and astrophysical plasmas by enabling
%%fast and explosive conversion of magnetic energy to plasma kinetic energy \cite{yamada_magnetic_2010}.
The HED plasmas is of 
general interest as a new platform for laboratory study of magnetic reconnection in high-beta regimes.
Previous simulations of reconnection in HED plasmas in which reconnection is driven by plasma flows \cite{fox_fast_2011,fox_magnetic_2012}
have recently been extended to demonstrate that
the magnetic field inflow can also be driven solely by the Nernst effect \cite{joglekar_magnetic_2014}
in a high-beta ($\beta_{p}\gg 1$) and semi-collisional ($\Omega \tau \approx 1.47 B[\mathrm{T}](T_{e}[\mathrm{keV}])^{3/2}/(Z n_{e}[10^{20} \mathrm{cm}^{-3}])\sim 1$) regime (hereafter denoted the Nernst regime), where $\Omega$ is the electron cyclotron frequency and $\tau$ is the electron collision time.
Interestingly, since the Nernst term is a pure advection term, it
vanishes at the magnetic field null point,  indicating
additional dissipation or decoupling mechanisms are required to allow for reconnection in this
regime.
 In simulations \cite{joglekar_magnetic_2014} it is
observed that momentum transport (given by off-diagonal components of the electron pressure tensor $\mathsf{\Pi}$) dominates
resistivity and provides the dissipation mechanism to break field lines in the reconnection layer.
The presence of such a momentum transport in the semi-collisional 
Nernst regime is highly interesting; for one, 
momentum transport
is well-established in \textit{collisionless} particle-in-cell (PIC) simulations of
reconnection \cite{hesse_collisionless_2001,bessho_collisionless_2005}.
However, finding closures for the kinetic equation to predict the magnitude of the momentum transport
in collisionless regimes has proved challenging\cite{wang_comparison_2015}.

In this paper, we present fully kinetic particle-in-cell simulations and analytic theory to elucidate the mechanism of magnetic reconnection driven by the Nernst effect,
 which demonstrates the interaction between nonlocal transport, 
 momentum transport, and the dynamics of the magnetic field.  
 Momentum transport in the current sheet provides the
 out-of-plane electric field for reconnection, and 
 is fundamentally a nonlocal process.
 The momentum transport obtained in the simulations is shown to be due to the ``heat-flux-viscosity'' \cite{catto_drift_2004,liu_heat_2015}, which describes the momentum transport arising from gradients in the heat-flux. 
 A consequence is that in the Nernst regime, 
 the half-width of the reconnection layer must be of 
the order of a few  electron mean free paths ($\lambda_{\mathrm{mfp}}\approx 110\mu \mathrm{m} (T_{e}/\mathrm{1keV})^{2} /(Z n_{e}/10^{20} \mathrm{cm}^{-3})$), a potentially experimentally-observable prediction.
At the mean-free-path scale, in the magnetic cavity of the reconnection layer, the transport transitions from diffusive to free-streaming, requiring nonlocal analysis\cite{kho_nonlinear_1985,luciani_magnetic_1985,ridgers_magnetic_2008,joglekar_kinetic_2016}.
This indicates that a typical heat flux driven magnetic reconnection process will have two stages: (1) 
a progressive formation of a thin current sheet as the two opposing fields 
are compressed together until, (2) the shear width becomes of the order of several mean-free-paths.
The very thin width of the layer is consistent with 
strong nonlocal effects \cite{lancia_topology_2014} inside the reconnection layer.
The results indicate how nonlocal effects and momentum transport are readily coupled 
for magnetic field transport in HED plasmas; this has often been ignored in 
previous calculations\cite{kho_nonlinear_1985,luciani_magnetic_1985} which assumed an isotropic distribution function and ignore momentum transport.

This paper is organized as follows. In Sec. \ref{sec:gol} we show the generalized Ohm's law derived from the transport theory, including the full contribution of the anisotropic pressure tensor. In Sec. \ref{sec:scaling} we show a scaling analysis of the GOL in the Nernst regime. In Sec. \ref{sec:pic} we demonstrate the results of a 2D collisional particle-in-cell simulation for heat-flux-driven magnetic reconnection, and analyze the advection and diffusion of the magnetic fields using the derived GOL. We also confirm that the origin of the anisotropic pressure tensor is mainly from the heat flux viscosity, and that nonlocal effects are important inside the reconnection layer. In Sec \ref{sec:conclusion} we conclude with a summary. In Appendix \ref{sec:appendix}, we show the details of derivation of the GOL.
%and the scaling of the transport coefficients with respect to the collionality of the plasma.

\section{Generalized Ohm's law with pressure tensor}
\label{sec:gol}

The dynamics of the magnetic field, including both advection and diffusion effects,
follow from a generalized Ohm's law for the electric field, in combination with Faraday's law.  For simplicity, 
we assume that ions are immobile using an extension of the electron MHD (eMHD) picture. (The inclusion of ion dynamics does not change the qualitative picture.)
The standard GOL can be calculated from the first-order moment of the kinetic equation,
\begin{equation}
  \label{ohm1}
  \mathbf{E}=\frac{\mathbf{R}_{ei}}{n_{e}e}+\frac{\mathbf{j}\times \mathbf{B}}{n_{e}ec}-\frac{\nabla \cdot \left(p_{e}\mathsf{I}+\mathsf{\Pi}\right)}{n_{e} e},
\end{equation}
where we have ignored the inertial term proportional to the small electron mass.
\footnote{We will see later in the simulation results that inertial term is very small.}
Here $\mathbf{E}$ is the electric field, $n_{e}$ is the electron density, $T_{e}$ is the electron temperature, $p_{e}=n_{e}T_{e}$, and $\mathsf{\Pi}$ is the traceless pressure tensor. $\mathbf{R}_{ei}$ is the friction force from electron-ion collisions, which  includes not only the resistivity effect, but also includes the thermal force,
which can be shown to contain the Nernst effect.  
Separating the contributions to $\mathbf{R}_{ei}$ in this manner, we derive a new GOL based on the the representation
\begin{equation}
f=f_{0}+\mathbf{f}_{1}\cdot \mathbf{v}/v+\underline{\underline{\mathsf{f}_{2}}}:\mathbf{v}\mathbf{v}/v^{2},
\end{equation}
and solve the kinetic equation for $\mathbf{f}_{1}$ following the steps in Appendix \ref{sec:appendix}.
Note that in the traditional derivation of the local transport theory\cite{braginskii_transport_1965,epperlein_plasma_1986}, contributions of $\mathsf{f}_{2}$ are ignored. But in the nonlocal regime\cite{bell_elecron_1981}, the transport model becomes non-perturbative and  the contribution from $\underline{\underline{\mathsf{f}_{2}}}$ can be of the same order as the other terms. Here we incorporate the contributions from $\mathsf{f}_{2}$  associated with the viscosity, which can be regarded as a first-order nonlocal transport effect in the GOL.

The new GOL can be written as,
\begin{align}
\label{ohm}
  \mathbf{E}=\frac{\underline{\underline{\alpha}}\cdot\mathbf{j}}{n_{e}^{2}e^{2}}+\frac{\mathbf{j}\times \mathbf{B}}{n_{e}ec}&-\frac{\nabla \cdot \left(p_{e}\mathsf{I}+\mathsf{\Pi}\right)}{n_{e} e}-\frac{\underline{\underline{\beta}}\cdot\nabla \cdot \left(T_{e}\mathsf{I}+\frac{2}{5}\mathsf{\Pi}/n_{e}\right)}{e},
\end{align}
where $\underline{\underline{\alpha}}$ is the resistivity tensor, which can be expressed as 
% which can be expressed using a dimensionless parameter, 
 $\underline{\underline\alpha}=\left(m_{e}n_{e}/\tau\right)\underline{\underline\alpha}^{c}$, where $\underline{\underline\alpha}^c$ is the dimensionless resistivity  (In this paper the superscript ``c'' denotes the dimensionless prefactor to the transport coefficients, which
 are generally functions of $\Omega \tau$ and the ion charge $Z$), and $\tau$ is the mean electron-ion collision time\cite{epperlein_plasma_1986},
\begin{equation}
  \tau=\frac{3\sqrt{m_{e}}T_{e}^{3/2}}{\sqrt{2\pi}n_{i}Z^{2}e^{4}\ln\Lambda}.
\end{equation} 
where $m_{e}$ is the electron mass, $n_{i}$ is the ion density, and $\ln\Lambda$ is the Coulomb logarithm.

In addition, $\underline{\underline{\beta}}$ is the thermoelectric tensor. The result of $\underline{\underline\alpha}^c$ and $\underline{\underline{\beta}}$ as functions of $Z$ and $\Omega\tau$ can be found in \cite{epperlein_plasma_1986}. The Nernst effect is contained in the off-diagonal terms of $\underline{\underline{\beta}}\cdot\nabla T_{e}$.
Note that the anisotropic pressure tensor, $\mathsf{\Pi}$, appears a second time in the thermoelectric term, which is different from Eq. (\ref{ohm1}).
We can see that like the resistivity, the pressure tensor introduces an additional dissipation effect in the GOL that can break field line and cause reconnection.

In a collisional plasma, the anisotropic pressure tensor, $\mathsf{\Pi}$, which describes the momentum transport in plasma, can be calculated from the viscosity in the transport theory. In addition to the standard flow viscosity which comes from the shear and compression of the flow, in \cite{liu_heat_2015} we showed the heat-flux viscosity (HFV) has a form analogous to the flow viscosity but with the replacement of
the plasma flow with the Nernst velocity. The anisotropic pressure tensor can then be calculated by combining the two viscosity effects,
\begin{equation}
\mathsf{\Pi}=\underline{\underline{\eta}}:\nabla \mathbf{v}_{e}+\underline{\underline{\mu}}:\nabla \mathbf{v}_{Ne},
\end{equation}
where $\mathbf{v}_{e}$ is the electron flow velocity, and $\mathbf{v}_{Ne}=2\mathbf{q}_{e}/(5n_{e}T_{e})$  is the electron Nernst velocity, where $\mathbf{q}_{e}$ is the electron heat flux \cite{haines_heat_1986}. $\underline{\underline{\eta}}$ and $\underline{\underline{\mu}}$ are the viscosity coefficients for 
particle-flow viscosity (PFV) and heat-flux viscosity (HFV), respectively. 
These coefficients are proportional to the electron collision time, which indicates that momentum transport matters when the mean free path is comparable to the gradient length scale. 
It can be regarded as a first-order nonlocal transport effect.
The picture of the HFV can also be understood by considering that an electron heat flux consists of counter-flowing 
electron populations at high and low energies; then, given that plasma viscosity decreases with collisionality, the high energy electrons
make a larger contribution to momentum transport than the low energy electrons, giving a net viscosity effect.
%
%We can now use the GOL to analyze the physics picture of the magnetic reconnection driven by laser heating. It is well-known the upstream magnetic fields can be strongly advected by the electron heat flux through Nernst effect\cite{haines_heat_1986,joglekar_magnetic_2014,joglekar_nernst_2015}.  We now investigate various terms in GOL and compare the relative amplitudes of them.%, in order to gain some insights of the physics picture of the heat flux driven magnetic reconnection.
%Note that we assume the ions are stationary  in order to isolate the electron transport effects.

%In our model the electron flow can be caused either by the pressure gradient or by the shear of the magnetic field. However, since the ions are immobile, electron flow driven by $\nabla p_{e}$ will be largely stopped by Coulomb force, so the current $j\sim \nabla B$ is the major part of the flow. The electron heat flux on the other hand can be driven by $\nabla T_{e}$ since it will not cause charge separation.

\section{Scaling analysis of the generalized Ohm's law}
\label{sec:scaling}

\begin{figure}[h]
  \centering
  \includegraphics[width=6cm]{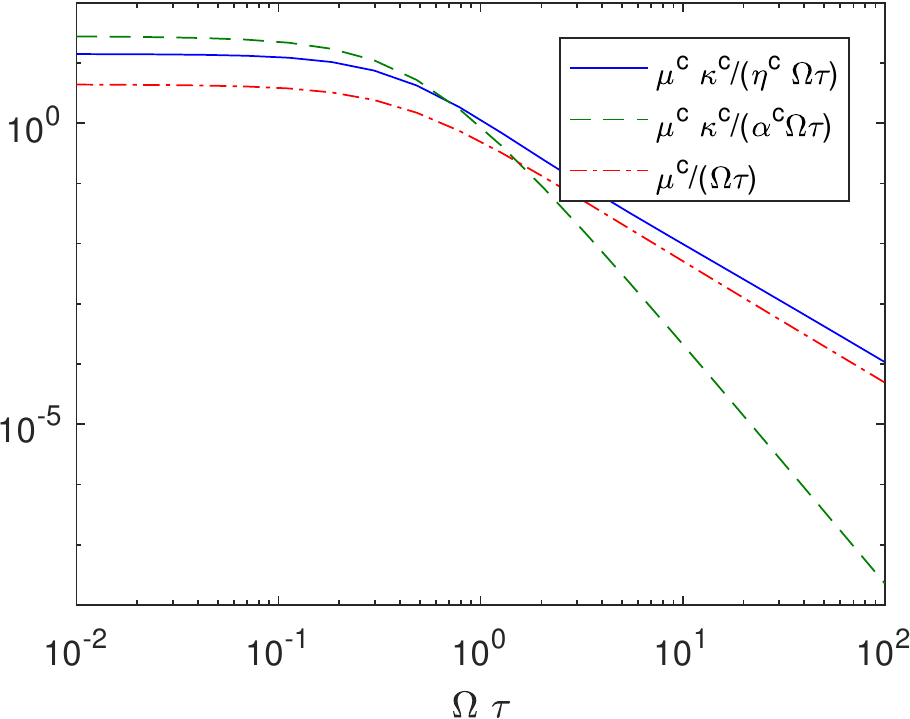}
  \caption{\label{fig:coefficients}The coefficients as functions of $\Omega\tau$ in the dimensionless analysis of the GOL.}
\end{figure}

We now provide a scaling analysis to compare the magnitudes of the terms in the GOL in
the Nernst regime to motivate the analysis of particle-in-cell simulations to follow.
We first estimate the ratio of contributions of HFV and PFV to dissipation in the reconnection layer:
%in our case can be simplified as,
\begin{equation}
  R_{1}=\frac{\mu \nabla \left(\kappa \nabla T_{e}/p_{e}\right)}{\eta \nabla \left(j_{\perp}/n_{e}e\right)}=\frac{\mu^{c}\kappa^{c}}{\eta^{c}}\frac{1}{\Omega\tau}\frac{\lambda_{\mathrm{mfp}}^{2}}{d_{e}^{2}}\frac{L_{S}}{L_{T}},
\end{equation}
where $j_{\perp}$ is the perpendicular current, $\kappa=\left(n_{e}T_{e}\tau\right)\kappa^{c}$ is the heat transfer coefficient,
$d_{e}=c/\omega_{pe}$ is the electron skin depth, and $\lambda_{\mathrm{mfp}}=v_{Te}\tau$ is the electron mean free path . Note that $\lambda_{mfp}^{2}/d_{e}^{2}=\beta_{p} (\Omega\tau)^{2}/2$, where $\beta_{p}$ is the plasma beta associated with the thermal energy. Here $L_{S}$ is the scale length of the magnetic shear, which is approximately the half-width of the reconnection layer, and $L_{T}$ is the scale length of temperature gradient driving
the reconnection.
It is interesting that this ratio is similar to $H_{N}$ of \cite{joglekar_magnetic_2014}, which describes a condition for the Nernst regime.
As shown in Fig. \ref{fig:coefficients}, the ratio of the prefactor, $\mu^{c}\kappa^{c}/(\eta^{c}\Omega\tau)$,
stays near 14 for $\Omega\tau \lesssim 0.1$, and then falls as $(\Omega\tau)^{-2}$ for $\Omega\tau\gg 1$. In the semi-collisional regime ($\Omega\tau\sim 1$), this demonstrates that if the plasma temperature is 
sufficiently high, such that $\lambda_{\mathrm{mfp}}\gg d_{e}$, then the HFV dominates the PFV in maintaining the pressure tensor.

We next compare the HFV and the resistivity in the GOL:
\begin{align}
  R_{2}=\frac{\nabla \Pi}{\alpha j_{\perp}}=\frac{\mu^{c}\kappa^{c}}{\alpha^{c}}\frac{1}{\Omega\tau}\frac{\lambda_{\mathrm{mfp}}^{4}}{L_{S} L_{T} d_{e}^{2}}.
\end{align}
The resulting prefactor, $\mu^{c}\kappa^{c}/\left(\alpha^{c}\Omega\tau\right)$, stays near 25 for $\Omega\tau \lesssim 0.1$, and falls as $(\Omega\tau)^{-4}$ asymptotically for $\Omega\tau\gg 1$.  If $L_{S}$ and $L_{T}$ are not too large, which is a necessary condition for a strong Nernst effect in the upstream, and again $\lambda_{\mathrm{mfp}}\gg d_{e}$, the pressure tensor contribution then dominates the resistivity.
The factor $\lambda_{\mathrm{mfp}}^{2}/(L_{S} L_{T})$ implies that the HFV enters as a nonlocal effect, where the $\lambda_{\mathrm{mfp}}$ is not ignorable compared to global length scales.
Note that  it has been shown that in the collisionless limit pressure tensor can balance the electric field in the reconnection layer and drive fast reconnection. Here we shown that even in semi-collisional regime this scenario is also possible.

Finally, we examine the balance between the HFV and the Nernst-driven inflow in the 
reconnection layer, as would be required to obtain a steady reconnection rate. Note that both terms depends on the heat flux $q_{e}$, so any flux-limitation effect due to the nonlocal transport will cancel. We obtain
\begin{equation}
  R_{3}=\frac{\nabla \Pi}{v_{N}B/c}=\frac{\mu^{c}}{\Omega\tau}\frac{\lambda_{\mathrm{mfp}}^{2}}{L_{S}^{2}}.
  \label{EqR3}
\end{equation}
The resulting prefactor, $\mu^{c}/(\Omega\tau)$, as shown in Fig. \ref{fig:coefficients}, has a similar trend to the previous two, which stays near 4.3 for $\Omega\tau \lesssim 0.1$, and goes as $(\Omega\tau)^{-2}$ asymptotically for $\Omega\tau\gg 1$.  This indicates that in a quasi-steady state, where the electric field in the upstream region and in the reconnection layer reach a balance, $L_{S}$ must be of the order of a few $\lambda_{\mathrm{mfp}}$, and that $L_{S}$ must decrease when $\Omega\tau$ increases.  This analysis indicates that in magnetic reconnection in Nernst regime, a thinning of the reconnection layer and a magnetic field compression\cite{fox_fast_2011,fox_magnetic_2012} can happen until $L_{S}$ becomes comparable to $\lambda_{\mathrm{mfp}}$.
The scaling of $R_{3}$ with mean-free-path shows that the HFV is manifestly a nonlocal effect.
\footnote{The result of $R_{3}$ indicates that for strongly magnetized case ($\Omega\tau\gg 1$) the layer width may be much smaller than $\lambda_{\mathrm{mfp}}$. However, we found this is not true because the \emph{additional} nonlocal transport effect, which will be discussed in Sec. \ref{sec:pic}, will introduce additional corrections to the amplitudes of the two terms.}.
In addition, given that $\lambda_{\mathrm{mfp}}^{2}$ is proportional to $\beta_{p}$, the occurrence of significant reconnection is related to large $\beta_{p}$.

%The above analysis suggests the following physics picture of the heat flux driven magnetic reconnection: 
%first the magnetic field 
%is advected by the Nernst effect from the upstream to the reconnection layer. 
%At early times when $L_{S}$ is still much larger than $\lambda_{\mathrm{mfp}}$,
%the HFV is small and the electric field at the reconnection layer remains smaller than in the upstream.  This results in
%a progressive thinning of the reconnection layer ($L_{S}$ decreases) and magnetic field compression 
%\cite{fox_fast_2011,fox_magnetic_2012}. When $L_{S}$ reaches a few  $\lambda_{\mathrm{mfp}}$, the contribution from 
%the pressure tensor in the reconnection layer becomes comparable to Nernst advection from the the upstream
%and a steady reconnection rate is reached.
%At this time the compression stops.
%The system then enters a quasi-steady state, that the field is continuously advected to the reconnection layer through the Nernst effect 
%and where they get reconnected through the pressure tensor effect.

\section{Particle-in-cell simulation}
\label{sec:pic}
We now directly study these processes in a collisional particle-in-cell (PIC) simulation using PSC \cite{germaschewski_plasma_2016}.
We simulate a 2D $x-z$ plane. 
The profile of density and temperature are 
 initialized to be uniform, with $n_{e}=n_{e0}$, $T_{e}=T_{e0}$. During the simulation, the plasma is heated in two semi-circular regions located at the center of the two $z$ boundaries to simulate the laser heating. In order to maintain a stationary temperature profile, we also artificially cool the plasma at the two $x$ boundaries to extract energy from the system. The magnetic field is initialized encircling the two hot spots 
 in a similar manner to previous simulations\cite{fox_fast_2011,joglekar_magnetic_2014}, with a peak value $B_{0}$. In the simulation we fix the ions to isolate the electron physics and simplify the analysis \footnote{Simulations with moving ions show nearly identical results.}.
 
 The parameters we used are as follows: $\Omega\tau=1.2$, $\lambda_{\mathrm{mfp}}/d_{e}=14$. The box size is $320 d_{e}\times 320 d_{e}$ (about $24 \lambda_{\mathrm{mfp}} \times24 \lambda_{\mathrm{mfp}}$).
We make a brief note on the determination of the parameters for our explicit particle-in-cell simulations.
The results of the PIC simulations can be applied and scaled to match a family of physical systems which match the relevant dimensionless parameters.
In the case of the Nernst problem considered here, the relevant parameters are: $\lambda_{\mathrm{mfp}}/d_{e}$, $\Omega\tau$, and $L/\lambda_{\mathrm{mfp}}$ (Other parameters which have appeared in the literature, such as $\beta_p$,  can be derived as combinations of these. In our simulation, the initial $\beta_{p}$ is 278).
Now, we note that if we choose, $T_{e}$ = 0.5 keV, $n_{e}=1.35\times 10^{21} \mathrm{cm}^{-3}$ and $B = 31$T, we will match the dimensionless parameters above.  These values are also close to current HEDP experiments.
To achieve these dimensionless parameters, while allowing an efficient explicit particle-in-cell simulation, we run at reduced speed of light $m_{e}c^{2}/T_{e} = 200$, rather than 1000 as is usually associated with $T_{e} = 0.5 $keV.
We note this type of compromise in the choice  of plasma parameters is common practice  in particle-in-cell simulations of magnetic reconnection phenomena\cite{daughton_influence_2009,fox_fast_2011} . While this ratio of scales is compressed, the parameter is well-matched in regime, and provides a basis for a convergence study.
In the case of the physics of the Nernst effect studied here, we notice that for the parameters chosen, the condition $v_N \ll c$ is satisfied, so the compression of this parameter is not expected to have significant physical consequences.
More directly, none of the ratios discussed in Sec. \ref{sec:scaling} depends on $m_{e}c^{2}/T_{e}$.
Finally, we conducted a convergence test which found that the primary results are converged with respect to this parameter.
The parameter does affect the population of electrons which are relativistic, and the scale separation of electrostatic and electromagnetic phenomena.

\begin{figure}[b]
	\centering
	\includegraphics[width=8cm]{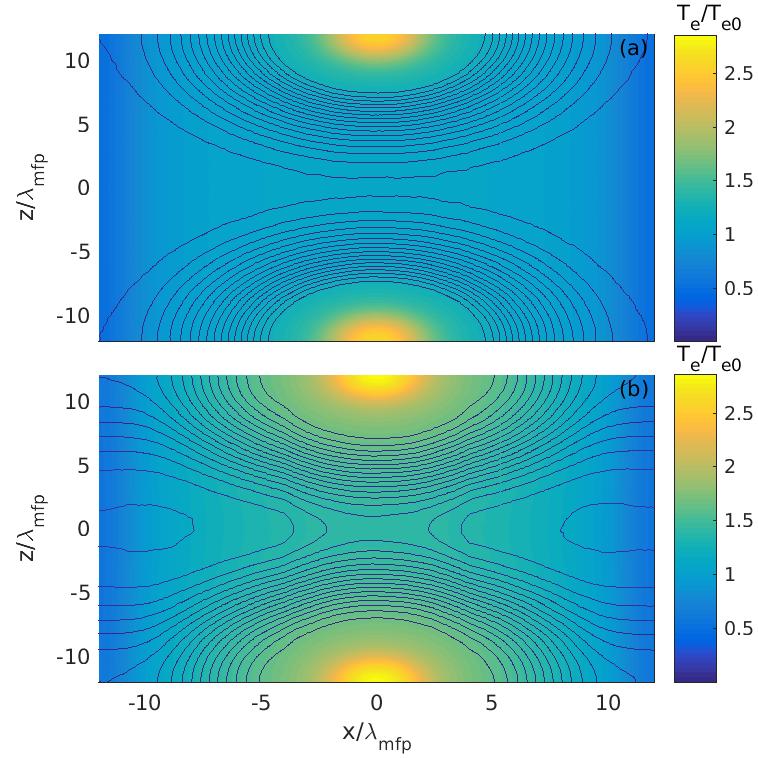}
	\caption{\label{fig:grid} The temperature profile (normalized to $T_{e0}$) and the magnetic flux contours at $t=10\tau$ (a) and $t=120\tau$ (b). $\lambda_{\mathrm{mfp}}$ is the mean free path calculated from initial density and temperature.}
\end{figure}

The temperature profile (normalized to the initial temperature $T_{e0}$) and the magnetic flux contours at $t=10\tau$ and $t=120\tau$ are shown in Fig. \ref{fig:grid}, which demonstrates that over time the magnetic flux is advected toward the mid-plane and is reconnected there through an X-point. We note that in the simulation we 
observed a collisional-Weibel instability\cite{epperlein_comparison_1985,thomas_rapid_2009} which causes ripples of the existing magnetic field and generates new magnetic field in the heating region.
To focus on the reconnection
within parameters easily accessed by a PIC simulation,
we force the magnetic field in the heating region to be zero, in order to suppress the instability
\footnote{The strong collisional-Weibel instability is partly due to the small-sized box in our PIC simulation due to the limitation of computation power. In a larger box the instability is expected to be much weaker thanks to the less focused heating region and smaller heat flux.}.

\begin{figure}[b]
	\centering
	\includegraphics[width=7.8cm]{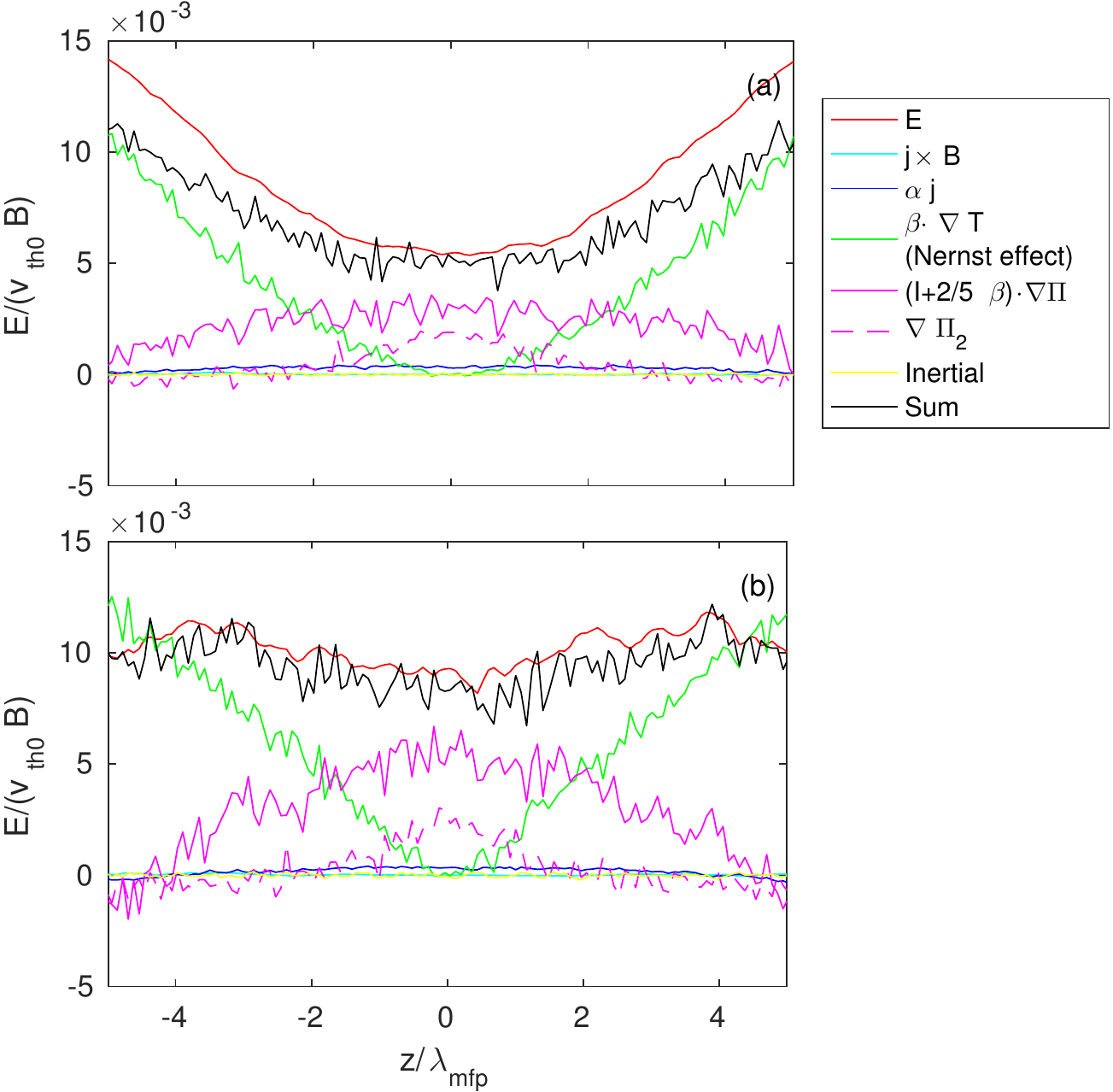}
	\caption{\label{fig:ohm} Contributions to the GOL for a cut across the reconnection layer near $x=0$ for (a) before reconnection ($t=10\tau$), (b) during steady reconnection ($t=120\tau$).  The red line is $E_{y}$ obtained from PIC simulation, and the black line shows the sum of the terms on the various contributing terms.  $E$ fields are measured in terms of $B_{0} v_{th0}$, where $B_{0}$ is the initial peak magnetic field in the simulation and $v_{\mathrm{th0}}=\sqrt{T_{e0}/m_e}$.}
\end{figure}

Fig. \ref{fig:ohm} shows the out-of-plane electric field, $E_{y}$, measured from the PIC simulation near $x=0$, and all the terms in the GOL calculated by taking moments of the distribution function obtained from the simulation. The sum of the terms comprising the right-hand-side of the GOL are in good agreement with $E_{y}$. We can see that at the early chosen time,
when $L_{S}$ is much larger than $\lambda_{\mathrm{mfp}}$, $E_y$ in the current sheet is 
smaller than the upstream, which results in subsequent compression of the upstream magnetic field.  At a later time,  the reconnection
has reached a steady state, where the electric field 
in the current sheet and upstream are comparable, mainly due to the increase of the pressure tensor term,
which dominates the GOL in the reconnection layer.
The magnetic fields are strongly reconnected in the center due to the dissipation from $\nabla\cdot\mathsf{\Pi}$. 
The pressure tensor term, though much larger, has a similar profile to the resistivity term. It thus can be regarded as an effective resistivity arising from the viscosity effect.
The $E_{y}$ in the reconnection layer in the quasisteady state is about $0.1 V_{N}$ ($V_{N}$ is the maximum Nernst velocity in the upstream), which is in agreement with the result in \cite{joglekar_magnetic_2014}.
As mentioned earlier, the contribution of the inertial term is negligible in this regime.
We note that, in order to show that the GOL is satisfied, we also include the contribution of the second order term in the expansion of $\mathsf{f}_{2}$, which is shown as the dashed purple line in Fig. \ref{fig:ohm}.
%In the steady state the electric field in the reconnection layer is about \(0.01 E_{D}\), where \(E_{D}\) is the Dreicer electric field. This indicates that the energization of the electron due to the electric field acceleration is not significant, and the electron distribution function is still close to Maxwellian.

\begin{figure}[b]
	\centering
	%\setbox1=\hbox{\includegraphics[height=2cm]{example-image-b}}
	\includegraphics[width=8cm]{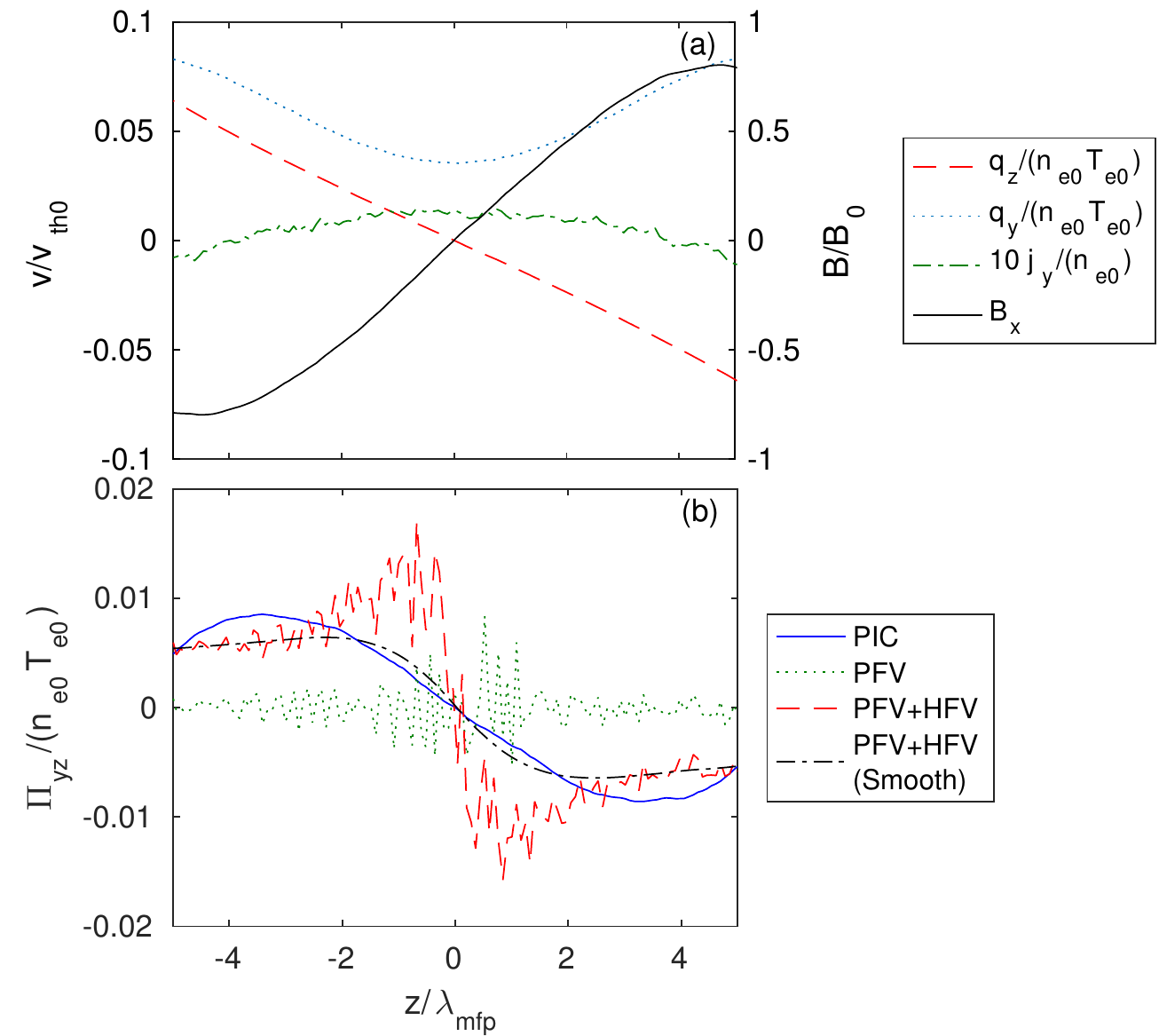}
	\caption{\label{fig:pressure}(a) Profiles of $q$, $j$ and $B$ in a cut across the reconnection layer near $x=0$ at $t=120\tau$. 
		(b) $\Pi_{yz}$ obtained from the PIC simulation, calculated from PFV and HFV using the quantities in the above plot, and a smoothed result applying the nonlocal kernel of Eq. (\ref{smooth}).}
\end{figure}

The simulation diagnostic output includes the direct momentum transport ($\mathsf{\Pi}$) and direct electron-ion collisional momentum transfer ($\mathbf{R}_{ei}$). We find that near the X-point $\mathbf{R}_{ei}$ and $\nabla\cdot\mathsf{\Pi}$ make approximately equal contributions. Recall that in the formulation of the Ohm's law used here, some of the direct momentum transfer ($\mathbf{R}_{ei}$) comes from the thermal force and appears as proportional to $\nabla\cdot\mathsf{\Pi}$.  The root of this is shown in Fig. \ref{fig:pressure}, where we show the structure of the current and heat flux.  We observe a finite out-of-plane heat flux ($q_{y}$) driven in the reconnection layer.  This heat flux cannot be the Righi-Luduc effect\cite{kho_nonlinear_1985}, since the magnetic field  crosses through zero in the layer.  Instead this heat flux is driven as a result of nonlocal effects.  Thermal-force-like friction on this heat flux leads to additional momentum transfer in the reconnection layer, which supports part of the reconnection electric field.
%This gives another picture of the origin of the heat flux viscosity effect, and clearly indicates it is a non-local effect.

Some recent laser-driven reconnection experiments have reported significant electron energization\cite{dong_plasmoid_2012,zhong_relativistic_2016}. It is therefore interesting to determine if a Nernst-driven reconnection can energize significant numbers of particles.  However, the results presented here do not show significant electron energization; typically the distribution functions remain close to Maxwellian and have not pulled out significant tails.
This can be understood from a simple estimation in the Nernst regime. Given that $E\sim v_N B$, there is $E/E_D\sim (v_N / v_{th}) (\Omega\tau)^{-1}$, where $E_{D}$ is the Dreicer electric field\cite{dreicer_electron_1959}.
Typically the former quantity is limited to be a fraction of 1, and the Nernst regime is semi-collisional ($\Omega\tau\sim 1$), yielding $E/E_{D}< 1$.  
Indeed, the simulation output shows $E/E_D\sim 0.01$. In addition, given that the scale length of the outflow region is much larger than $\lambda_{\mathrm{mfp}}$, one can expect strong thermal heating in the outflow direction, and energy from reconnection is converted to thermal energy through collisions.

We now investigate the origin of the pressure tensor that contributes to the GOL in the reconnection layer. 
To compare PFV and HFV, we calculate the contribution from both types of viscosity with $\Pi_{yz}$ obtained directly from the PIC simulation in Fig. \ref{fig:pressure}. The HFV is found to dominate the PFV, consistent with our previous scaling analysis. 
%The major contribution to the momentum 
%transport arises from the component 
%% to the $y$ component of $\nabla\cdot\mathsf{\Pi}$ comes from the component
% $\Pi_{yz}$, which can be generated through both PFV and HFV. 
 %Among the HFV effects,  
 As shown in Fig. \ref{fig:pressure}, $\Pi_{yz}$ can be generated through both 
 the compression term $\partial_{z}q_{z}$,
%\begin{align}
%\label{f2}
%  \frac{\partial \mathsf{f}_2}{\partial t}+\left(v\nabla \mathbf{f}_1-\frac{v}{3}\mathsf{I} \nabla\cdot \mathbf{f}_1\right)-\left[v\frac{\partial}{\partial v}\frac{\mathbf{E}e}{m}\frac{\mathbf{f}_1}{v}-\frac{v}{3}\left(\frac{\mathbf{E}e}{m}\cdot\frac{\partial}{\partial v}\frac{\mathbf{f}_1}{v}\right)\mathsf{I}\right]\nonumber\\
%  +2\mathbf{\Omega}\times \mathbf{f}_2+\frac{3}{7}v\nabla\cdot \mathbf{f}_3-\frac{3}{7v^{4}}\frac{\partial}{\partial v}\left(v^4 \frac{\mathbf{E}e}{m}\cdot \mathbf{f}_3\right)=C_2(\mathsf{f}_2),
%\end{align}
and the shear term $\partial_{z}q_{y}$. The two contributions are found to be of similar magnitude.

Interestingly, we find that although the magnitudes of the pressure tensor in the simulation
and calculated from HFV are similar, the gradients inside the reconnection layer are significantly different,
with the HFV theory over-predicting the resulting $\nabla\cdot\mathsf{\Pi}$.
This can be attributed to \textit{additional} nonlocal effects in the heat flux viscosity, which is related to the $\mathsf{f}_{3}$ term in Eq. (4) in \cite{liu_heat_2015}.
%, which is ignored in the Braginskii's transport theory, and the nonlocal effect given by the magnetic shear. 
Imagine that if $\Pi_{yz}$ calculated from PFV and HFV has a very large gradient in the reconnection layer, the gradient can then give rise to $\mathsf{f}_{3}$, which in return affects $\mathsf{f}_{2}$ and makes $\mathsf{f}_{2}$ profile smoother. 
To get a qualitative picture of this effect, we apply a nonlocal operator,
 to simulate the effect of $\mathsf{f}_{3}$ to $\Pi_{yz}$,
\begin{equation}
\label{smooth}
  (1-\rho^{2}\nabla^{2})\Pi^{s}_{yz}=\Pi_{yz},
\end{equation}
where $\rho=1/\sqrt{1/\rho_{L}^{2}+1/\lambda_{\mathrm{mfp}}^{2}}$, $\rho_{L}$ is the Larmor radius.
Note that this operator is consistent with the Luciani-Mora-Virmont model\cite{luciani_nonlocal_1983} for nonlocal heat transport. The result of $\Pi^{s}_{yz}$ is shown in Fig. \ref{fig:pressure}, which shows much better agreement with the simulation result. This smoothed pressure tensor can then be plugged into Eq. (\ref{ohm}). In this manner, the pressure tensor term in the GOL already includes the nonlocal effect.
This is the same approach for the application of the Nernst effect, where the Nernst term in the GOL is calculated from a flux-limiter or a nonlocal heat conduction model\cite{lancia_topology_2014}.
To better study this effect quantitatively, 
a self-consistent closure of the nonlocal viscosity, like a Landau fluid model\cite{hammett_fluid_1990,snyder_landau_1997},
will be considered further in future work.

Finally, we have conducted multiple runs at various $\Omega\tau$ to further verify the theory. We find that as $\Omega\tau$ increases the equilibrium $L_{S}$ decreases and the reconnection layer becomes narrower.
As we vary the magnetic field strength from $\Omega \tau =$0.9 to 5.8, 
the width of the shear layer decreases from 3 to 1.2, measured in units of the 
local mean-free-path.
% The reconnection picture then converts to an eMHD Sweet-Parker-like reconnection. 
% Scaling analysis of this regime shows that the equilibrium $L_{S}$ is about one $\lambda_{\mathrm{mfp}}$, which is verified in the simulation results. 
 We also note that when $\Omega\tau<1$, the equilibrium $L_{S}$ can be larger than $\lambda_{\mathrm{mfp}}$, and consequently the 
\textit{additional} nonlocal effects are found to be smaller and the calculated profile of pressure tensor shows a better agreement with the simulation.

\section{Conclusions}
\label{sec:conclusion}
We have depicted the picture of heat-flux-driven magnetic reconnection in HED plasmas using a generalized 
Ohm's law in which we include the heat-flux viscosity as a nonlocal dissipation mechanism.
Our calculations show that the shear and compression of the heat flux gives rise to momentum transport, which
allows for reconnection in this regime.  We find that the balance of the GOL 
in the upstream and in the reconnection layer sets the characteristic width of the reconnection layer to be
of the order of several mean free paths. %, which indicates a two-step reconnection process.
%The narrow layer-width indicates significant non-local transport inside the reconnection layer.
These results show the important interplay between nonlocal transport effects and generation of anisotropic components to the distribution function.  

Common approaches to nonlocal transport based on Fokker-Planck simulation truncate the distribution function after the first order
\cite{kho_nonlinear_1985,luciani_magnetic_1985}. (For a recent review of Fokker-Planck simulation, including results from extension to higher order, see \cite{thomas_review_2012})
However, our results demonstrate how higher order anisotropic terms can be generated in magnetized plasmas. These effects manifest as momentum transport and, as shown here, make an important contribution to the evolution of the magnetic field via GOL.  The scaling of the off-diagonal terms (i.e. heat-flux viscosity) in comparison with the Nernst term (Eq.~\ref{EqR3})
documented here confirms that these effects enter at the same order as other nonlocal transport effects \cite{kho_nonlinear_1985},
calling for further study of the coupling of momentum transport to the dynamics of magnetic fields in nonlocal transport regimes.

One of the authors (Chang Liu) wants to thank John Krommes, Gregory W. Hammett and Eero Hirvijoki for valuable discussions. The PIC simulations in this work were conducted on the Hopper and Cori supercomputers at the National Energy Research Scientific Computing Center, 
supported by the U. S. Department of Energy under Contract No. DE-AC02-05CH11231, 
and the Titan supercomputer at the Oak Ridge Leadership Computing Facility at the Oak Ridge National Laboratory, supported by the Office of Science of the U.S. Department of Energy under Contract No. DE-AC05-00OR22725. This work was supported by the U.S.~Department of Energy under Contract No. DE-SC0008655 and No. DE-SC0010621. A. S. Joglekar would like to acknowledge support from DE-NA0002953 and NSF ACI-1339893.
%add eero and john greg here
\bibliography{Nernst}

\begin{thebibliography}{47}%
\makeatletter
\providecommand \@ifxundefined [1]{%
 \@ifx{#1\undefined}
}%
\providecommand \@ifnum [1]{%
 \ifnum #1\expandafter \@firstoftwo
 \else \expandafter \@secondoftwo
 \fi
}%
\providecommand \@ifx [1]{%
 \ifx #1\expandafter \@firstoftwo
 \else \expandafter \@secondoftwo
 \fi
}%
\providecommand \natexlab [1]{#1}%
\providecommand \enquote  [1]{``#1''}%
\providecommand \bibnamefont  [1]{#1}%
\providecommand \bibfnamefont [1]{#1}%
\providecommand \citenamefont [1]{#1}%
\providecommand \href@noop [0]{\@secondoftwo}%
\providecommand \href [0]{\begingroup \@sanitize@url \@href}%
\providecommand \@href[1]{\@@startlink{#1}\@@href}%
\providecommand \@@href[1]{\endgroup#1\@@endlink}%
\providecommand \@sanitize@url [0]{\catcode `\\12\catcode `\$12\catcode
  `\&12\catcode `\#12\catcode `\^12\catcode `\_12\catcode `\%12\relax}%
\providecommand \@@startlink[1]{}%
\providecommand \@@endlink[0]{}%
\providecommand \url  [0]{\begingroup\@sanitize@url \@url }%
\providecommand \@url [1]{\endgroup\@href {#1}{\urlprefix }}%
\providecommand \urlprefix  [0]{URL }%
\providecommand \Eprint [0]{\href }%
\providecommand \doibase [0]{http://dx.doi.org/}%
\providecommand \selectlanguage [0]{\@gobble}%
\providecommand \bibinfo  [0]{\@secondoftwo}%
\providecommand \bibfield  [0]{\@secondoftwo}%
\providecommand \translation [1]{[#1]}%
\providecommand \BibitemOpen [0]{}%
\providecommand \bibitemStop [0]{}%
\providecommand \bibitemNoStop [0]{.\EOS\space}%
\providecommand \EOS [0]{\spacefactor3000\relax}%
\providecommand \BibitemShut  [1]{\csname bibitem#1\endcsname}%
\let\auto@bib@innerbib\@empty
%</preamble>
\bibitem [{\citenamefont {Stamper}\ \emph {et~al.}(1971)\citenamefont
  {Stamper}, \citenamefont {Papadopoulos}, \citenamefont {Sudan}, \citenamefont
  {Dean}, \citenamefont {McLean},\ and\ \citenamefont
  {Dawson}}]{stamper_spontaneous_1971}%
  \BibitemOpen
  \bibfield  {author} {\bibinfo {author} {\bibfnamefont {J.~A.}\ \bibnamefont
  {Stamper}}, \bibinfo {author} {\bibfnamefont {K.}~\bibnamefont
  {Papadopoulos}}, \bibinfo {author} {\bibfnamefont {R.~N.}\ \bibnamefont
  {Sudan}}, \bibinfo {author} {\bibfnamefont {S.~O.}\ \bibnamefont {Dean}},
  \bibinfo {author} {\bibfnamefont {E.~A.}\ \bibnamefont {McLean}}, \ and\
  \bibinfo {author} {\bibfnamefont {J.~M.}\ \bibnamefont {Dawson}},\ }\href
  {\doibase 10.1103/PhysRevLett.26.1012} {\bibfield  {journal} {\bibinfo
  {journal} {Phys. Rev. Lett.}\ }\textbf {\bibinfo {volume} {26}},\ \bibinfo
  {pages} {1012} (\bibinfo {year} {1971})}\BibitemShut {NoStop}%
\bibitem [{\citenamefont {Rygg}\ \emph {et~al.}(2008)\citenamefont {Rygg},
  \citenamefont {S{\'e}guin}, \citenamefont {Li}, \citenamefont {Frenje},
  \citenamefont {Manuel}, \citenamefont {Petrasso}, \citenamefont {Betti},
  \citenamefont {Delettrez}, \citenamefont {Gotchev}, \citenamefont {Knauer},
  \citenamefont {Meyerhofer}, \citenamefont {Marshall}, \citenamefont
  {Stoeckl},\ and\ \citenamefont {Theobald}}]{rygg_proton_2008}%
  \BibitemOpen
  \bibfield  {author} {\bibinfo {author} {\bibfnamefont {J.~R.}\ \bibnamefont
  {Rygg}}, \bibinfo {author} {\bibfnamefont {F.~H.}\ \bibnamefont
  {S{\'e}guin}}, \bibinfo {author} {\bibfnamefont {C.~K.}\ \bibnamefont {Li}},
  \bibinfo {author} {\bibfnamefont {J.~A.}\ \bibnamefont {Frenje}}, \bibinfo
  {author} {\bibfnamefont {M.~J.-E.}\ \bibnamefont {Manuel}}, \bibinfo {author}
  {\bibfnamefont {R.~D.}\ \bibnamefont {Petrasso}}, \bibinfo {author}
  {\bibfnamefont {R.}~\bibnamefont {Betti}}, \bibinfo {author} {\bibfnamefont
  {J.~A.}\ \bibnamefont {Delettrez}}, \bibinfo {author} {\bibfnamefont {O.~V.}\
  \bibnamefont {Gotchev}}, \bibinfo {author} {\bibfnamefont {J.~P.}\
  \bibnamefont {Knauer}}, \bibinfo {author} {\bibfnamefont {D.~D.}\
  \bibnamefont {Meyerhofer}}, \bibinfo {author} {\bibfnamefont {F.~J.}\
  \bibnamefont {Marshall}}, \bibinfo {author} {\bibfnamefont {C.}~\bibnamefont
  {Stoeckl}}, \ and\ \bibinfo {author} {\bibfnamefont {W.}~\bibnamefont
  {Theobald}},\ }\href {\doibase 10.1126/science.1152640} {\bibfield  {journal}
  {\bibinfo  {journal} {Science}\ }\textbf {\bibinfo {volume} {319}},\ \bibinfo
  {pages} {1223} (\bibinfo {year} {2008})}\BibitemShut {NoStop}%
\bibitem [{\citenamefont {Gao}\ \emph {et~al.}(2012)\citenamefont {Gao},
  \citenamefont {Nilson}, \citenamefont {Igumenschev}, \citenamefont {Hu},
  \citenamefont {Davies}, \citenamefont {Stoeckl}, \citenamefont {Haines},
  \citenamefont {Froula}, \citenamefont {Betti},\ and\ \citenamefont
  {Meyerhofer}}]{gao_magnetic_2012}%
  \BibitemOpen
  \bibfield  {author} {\bibinfo {author} {\bibfnamefont {L.}~\bibnamefont
  {Gao}}, \bibinfo {author} {\bibfnamefont {P.~M.}\ \bibnamefont {Nilson}},
  \bibinfo {author} {\bibfnamefont {I.~V.}\ \bibnamefont {Igumenschev}},
  \bibinfo {author} {\bibfnamefont {S.~X.}\ \bibnamefont {Hu}}, \bibinfo
  {author} {\bibfnamefont {J.~R.}\ \bibnamefont {Davies}}, \bibinfo {author}
  {\bibfnamefont {C.}~\bibnamefont {Stoeckl}}, \bibinfo {author} {\bibfnamefont
  {M.~G.}\ \bibnamefont {Haines}}, \bibinfo {author} {\bibfnamefont {D.~H.}\
  \bibnamefont {Froula}}, \bibinfo {author} {\bibfnamefont {R.}~\bibnamefont
  {Betti}}, \ and\ \bibinfo {author} {\bibfnamefont {D.~D.}\ \bibnamefont
  {Meyerhofer}},\ }\href {\doibase 10.1103/PhysRevLett.109.115001} {\bibfield
  {journal} {\bibinfo  {journal} {Phys. Rev. Lett.}\ }\textbf {\bibinfo
  {volume} {109}},\ \bibinfo {pages} {115001} (\bibinfo {year}
  {2012})}\BibitemShut {NoStop}%
\bibitem [{\citenamefont {Fox}\ \emph {et~al.}(2013)\citenamefont {Fox},
  \citenamefont {Fiksel}, \citenamefont {Bhattacharjee}, \citenamefont {Chang},
  \citenamefont {Germaschewski}, \citenamefont {Hu},\ and\ \citenamefont
  {Nilson}}]{fox_filamentation_2013}%
  \BibitemOpen
  \bibfield  {author} {\bibinfo {author} {\bibfnamefont {W.}~\bibnamefont
  {Fox}}, \bibinfo {author} {\bibfnamefont {G.}~\bibnamefont {Fiksel}},
  \bibinfo {author} {\bibfnamefont {A.}~\bibnamefont {Bhattacharjee}}, \bibinfo
  {author} {\bibfnamefont {P.-Y.}\ \bibnamefont {Chang}}, \bibinfo {author}
  {\bibfnamefont {K.}~\bibnamefont {Germaschewski}}, \bibinfo {author}
  {\bibfnamefont {S.~X.}\ \bibnamefont {Hu}}, \ and\ \bibinfo {author}
  {\bibfnamefont {P.~M.}\ \bibnamefont {Nilson}},\ }\href {\doibase
  10.1103/PhysRevLett.111.225002} {\bibfield  {journal} {\bibinfo  {journal}
  {Phys. Rev. Lett.}\ }\textbf {\bibinfo {volume} {111}},\ \bibinfo {pages}
  {225002} (\bibinfo {year} {2013})}\BibitemShut {NoStop}%
\bibitem [{\citenamefont {Huntington}\ \emph {et~al.}(2015)\citenamefont
  {Huntington}, \citenamefont {Fiuza}, \citenamefont {Ross}, \citenamefont
  {Zylstra}, \citenamefont {Drake}, \citenamefont {Froula}, \citenamefont
  {Gregori}, \citenamefont {Kugland}, \citenamefont {Kuranz}, \citenamefont
  {Levy}, \citenamefont {Li}, \citenamefont {Meinecke}, \citenamefont {Morita},
  \citenamefont {Petrasso}, \citenamefont {Plechaty}, \citenamefont
  {Remington}, \citenamefont {Ryutov}, \citenamefont {Sakawa}, \citenamefont
  {Spitkovsky}, \citenamefont {Takabe},\ and\ \citenamefont
  {Park}}]{huntington_observation_2015}%
  \BibitemOpen
  \bibfield  {author} {\bibinfo {author} {\bibfnamefont {C.~M.}\ \bibnamefont
  {Huntington}}, \bibinfo {author} {\bibfnamefont {F.}~\bibnamefont {Fiuza}},
  \bibinfo {author} {\bibfnamefont {J.~S.}\ \bibnamefont {Ross}}, \bibinfo
  {author} {\bibfnamefont {A.~B.}\ \bibnamefont {Zylstra}}, \bibinfo {author}
  {\bibfnamefont {R.~P.}\ \bibnamefont {Drake}}, \bibinfo {author}
  {\bibfnamefont {D.~H.}\ \bibnamefont {Froula}}, \bibinfo {author}
  {\bibfnamefont {G.}~\bibnamefont {Gregori}}, \bibinfo {author} {\bibfnamefont
  {N.~L.}\ \bibnamefont {Kugland}}, \bibinfo {author} {\bibfnamefont {C.~C.}\
  \bibnamefont {Kuranz}}, \bibinfo {author} {\bibfnamefont {M.~C.}\
  \bibnamefont {Levy}}, \bibinfo {author} {\bibfnamefont {C.~K.}\ \bibnamefont
  {Li}}, \bibinfo {author} {\bibfnamefont {J.}~\bibnamefont {Meinecke}},
  \bibinfo {author} {\bibfnamefont {T.}~\bibnamefont {Morita}}, \bibinfo
  {author} {\bibfnamefont {R.}~\bibnamefont {Petrasso}}, \bibinfo {author}
  {\bibfnamefont {C.}~\bibnamefont {Plechaty}}, \bibinfo {author}
  {\bibfnamefont {B.~A.}\ \bibnamefont {Remington}}, \bibinfo {author}
  {\bibfnamefont {D.~D.}\ \bibnamefont {Ryutov}}, \bibinfo {author}
  {\bibfnamefont {Y.}~\bibnamefont {Sakawa}}, \bibinfo {author} {\bibfnamefont
  {A.}~\bibnamefont {Spitkovsky}}, \bibinfo {author} {\bibfnamefont
  {H.}~\bibnamefont {Takabe}}, \ and\ \bibinfo {author} {\bibfnamefont {H.-S.}\
  \bibnamefont {Park}},\ }\href {\doibase 10.1038/nphys3178} {\bibfield
  {journal} {\bibinfo  {journal} {Nature Phys.}\ }\textbf {\bibinfo {volume}
  {11}},\ \bibinfo {pages} {173} (\bibinfo {year} {2015})}\BibitemShut
  {NoStop}%
\bibitem [{\citenamefont {Glenzer}\ \emph {et~al.}(1999)\citenamefont
  {Glenzer}, \citenamefont {Alley}, \citenamefont {Estabrook}, \citenamefont
  {Groot}, \citenamefont {Haines}, \citenamefont {Hammer}, \citenamefont
  {Jadaud}, \citenamefont {MacGowan}, \citenamefont {Moody}, \citenamefont
  {Rozmus}, \citenamefont {Suter}, \citenamefont {Weiland},\ and\ \citenamefont
  {Williams}}]{glenzer_thomson_1999}%
  \BibitemOpen
  \bibfield  {author} {\bibinfo {author} {\bibfnamefont {S.~H.}\ \bibnamefont
  {Glenzer}}, \bibinfo {author} {\bibfnamefont {W.~E.}\ \bibnamefont {Alley}},
  \bibinfo {author} {\bibfnamefont {K.~G.}\ \bibnamefont {Estabrook}}, \bibinfo
  {author} {\bibfnamefont {J.~S.~D.}\ \bibnamefont {Groot}}, \bibinfo {author}
  {\bibfnamefont {M.~G.}\ \bibnamefont {Haines}}, \bibinfo {author}
  {\bibfnamefont {J.~H.}\ \bibnamefont {Hammer}}, \bibinfo {author}
  {\bibfnamefont {J.-P.}\ \bibnamefont {Jadaud}}, \bibinfo {author}
  {\bibfnamefont {B.~J.}\ \bibnamefont {MacGowan}}, \bibinfo {author}
  {\bibfnamefont {J.~D.}\ \bibnamefont {Moody}}, \bibinfo {author}
  {\bibfnamefont {W.}~\bibnamefont {Rozmus}}, \bibinfo {author} {\bibfnamefont
  {L.~J.}\ \bibnamefont {Suter}}, \bibinfo {author} {\bibfnamefont {T.~L.}\
  \bibnamefont {Weiland}}, \ and\ \bibinfo {author} {\bibfnamefont {E.~A.}\
  \bibnamefont {Williams}},\ }\href {\doibase 10.1063/1.873499} {\bibfield
  {journal} {\bibinfo  {journal} {Phys. Plasmas}\ }\textbf {\bibinfo {volume}
  {6}},\ \bibinfo {pages} {2117} (\bibinfo {year} {1999})}\BibitemShut
  {NoStop}%
\bibitem [{\citenamefont {Slutz}\ \emph {et~al.}(2010)\citenamefont {Slutz},
  \citenamefont {Herrmann}, \citenamefont {Vesey}, \citenamefont {Sefkow},
  \citenamefont {Sinars}, \citenamefont {Rovang}, \citenamefont {Peterson},\
  and\ \citenamefont {Cuneo}}]{slutz_pulsed-power-driven_2010}%
  \BibitemOpen
  \bibfield  {author} {\bibinfo {author} {\bibfnamefont {S.~A.}\ \bibnamefont
  {Slutz}}, \bibinfo {author} {\bibfnamefont {M.~C.}\ \bibnamefont {Herrmann}},
  \bibinfo {author} {\bibfnamefont {R.~A.}\ \bibnamefont {Vesey}}, \bibinfo
  {author} {\bibfnamefont {A.~B.}\ \bibnamefont {Sefkow}}, \bibinfo {author}
  {\bibfnamefont {D.~B.}\ \bibnamefont {Sinars}}, \bibinfo {author}
  {\bibfnamefont {D.~C.}\ \bibnamefont {Rovang}}, \bibinfo {author}
  {\bibfnamefont {K.~J.}\ \bibnamefont {Peterson}}, \ and\ \bibinfo {author}
  {\bibfnamefont {M.~E.}\ \bibnamefont {Cuneo}},\ }\href {\doibase
  10.1063/1.3333505} {\bibfield  {journal} {\bibinfo  {journal} {Phys.
  Plasmas}\ }\textbf {\bibinfo {volume} {17}},\ \bibinfo {pages} {056303}
  (\bibinfo {year} {2010})}\BibitemShut {NoStop}%
\bibitem [{\citenamefont {Chang}\ \emph {et~al.}(2011)\citenamefont {Chang},
  \citenamefont {Fiksel}, \citenamefont {Hohenberger}, \citenamefont {Knauer},
  \citenamefont {Betti}, \citenamefont {Marshall}, \citenamefont {Meyerhofer},
  \citenamefont {S{\'e}guin},\ and\ \citenamefont
  {Petrasso}}]{chang_fusion_2011}%
  \BibitemOpen
  \bibfield  {author} {\bibinfo {author} {\bibfnamefont {P.~Y.}\ \bibnamefont
  {Chang}}, \bibinfo {author} {\bibfnamefont {G.}~\bibnamefont {Fiksel}},
  \bibinfo {author} {\bibfnamefont {M.}~\bibnamefont {Hohenberger}}, \bibinfo
  {author} {\bibfnamefont {J.~P.}\ \bibnamefont {Knauer}}, \bibinfo {author}
  {\bibfnamefont {R.}~\bibnamefont {Betti}}, \bibinfo {author} {\bibfnamefont
  {F.~J.}\ \bibnamefont {Marshall}}, \bibinfo {author} {\bibfnamefont {D.~D.}\
  \bibnamefont {Meyerhofer}}, \bibinfo {author} {\bibfnamefont {F.~H.}\
  \bibnamefont {S{\'e}guin}}, \ and\ \bibinfo {author} {\bibfnamefont {R.~D.}\
  \bibnamefont {Petrasso}},\ }\href {\doibase 10.1103/PhysRevLett.107.035006}
  {\bibfield  {journal} {\bibinfo  {journal} {Phys. Rev. Lett.}\ }\textbf
  {\bibinfo {volume} {107}},\ \bibinfo {pages} {035006} (\bibinfo {year}
  {2011})}\BibitemShut {NoStop}%
\bibitem [{\citenamefont {Haines}(1986)}]{haines_heat_1986}%
  \BibitemOpen
  \bibfield  {author} {\bibinfo {author} {\bibfnamefont {M.~G.}\ \bibnamefont
  {Haines}},\ }\href {\doibase 10.1088/0741-3335/28/11/007} {\bibfield
  {journal} {\bibinfo  {journal} {Plasma Phys. Control. Fusion}\ }\textbf
  {\bibinfo {volume} {28}},\ \bibinfo {pages} {1705} (\bibinfo {year}
  {1986})}\BibitemShut {NoStop}%
\bibitem [{\citenamefont {Kho}\ and\ \citenamefont
  {Haines}(1985)}]{kho_nonlinear_1985}%
  \BibitemOpen
  \bibfield  {author} {\bibinfo {author} {\bibfnamefont {T.~H.}\ \bibnamefont
  {Kho}}\ and\ \bibinfo {author} {\bibfnamefont {M.~G.}\ \bibnamefont
  {Haines}},\ }\href {\doibase 10.1103/PhysRevLett.55.825} {\bibfield
  {journal} {\bibinfo  {journal} {Phys. Rev. Lett.}\ }\textbf {\bibinfo
  {volume} {55}},\ \bibinfo {pages} {825} (\bibinfo {year} {1985})}\BibitemShut
  {NoStop}%
\bibitem [{\citenamefont {Ridgers}\ \emph {et~al.}(2008)\citenamefont
  {Ridgers}, \citenamefont {Kingham},\ and\ \citenamefont
  {Thomas}}]{ridgers_magnetic_2008}%
  \BibitemOpen
  \bibfield  {author} {\bibinfo {author} {\bibfnamefont {C.~P.}\ \bibnamefont
  {Ridgers}}, \bibinfo {author} {\bibfnamefont {R.~J.}\ \bibnamefont
  {Kingham}}, \ and\ \bibinfo {author} {\bibfnamefont {A.~G.~R.}\ \bibnamefont
  {Thomas}},\ }\href {\doibase 10.1103/PhysRevLett.100.075003} {\bibfield
  {journal} {\bibinfo  {journal} {Phys. Rev. Lett.}\ }\textbf {\bibinfo
  {volume} {100}},\ \bibinfo {pages} {075003} (\bibinfo {year}
  {2008})}\BibitemShut {NoStop}%
\bibitem [{\citenamefont {Froula}\ \emph {et~al.}(2007)\citenamefont {Froula},
  \citenamefont {Ross}, \citenamefont {Pollock}, \citenamefont {Davis},
  \citenamefont {James}, \citenamefont {Divol}, \citenamefont {Edwards},
  \citenamefont {Offenberger}, \citenamefont {Price}, \citenamefont {Town},
  \citenamefont {Tynan},\ and\ \citenamefont
  {Glenzer}}]{froula_quenching_2007}%
  \BibitemOpen
  \bibfield  {author} {\bibinfo {author} {\bibfnamefont {D.~H.}\ \bibnamefont
  {Froula}}, \bibinfo {author} {\bibfnamefont {J.~S.}\ \bibnamefont {Ross}},
  \bibinfo {author} {\bibfnamefont {B.~B.}\ \bibnamefont {Pollock}}, \bibinfo
  {author} {\bibfnamefont {P.}~\bibnamefont {Davis}}, \bibinfo {author}
  {\bibfnamefont {A.~N.}\ \bibnamefont {James}}, \bibinfo {author}
  {\bibfnamefont {L.}~\bibnamefont {Divol}}, \bibinfo {author} {\bibfnamefont
  {M.~J.}\ \bibnamefont {Edwards}}, \bibinfo {author} {\bibfnamefont {A.~A.}\
  \bibnamefont {Offenberger}}, \bibinfo {author} {\bibfnamefont
  {D.}~\bibnamefont {Price}}, \bibinfo {author} {\bibfnamefont {R.~P.~J.}\
  \bibnamefont {Town}}, \bibinfo {author} {\bibfnamefont {G.~R.}\ \bibnamefont
  {Tynan}}, \ and\ \bibinfo {author} {\bibfnamefont {S.~H.}\ \bibnamefont
  {Glenzer}},\ }\href {\doibase 10.1103/PhysRevLett.98.135001} {\bibfield
  {journal} {\bibinfo  {journal} {Phys. Rev. Lett.}\ }\textbf {\bibinfo
  {volume} {98}},\ \bibinfo {pages} {135001} (\bibinfo {year}
  {2007})}\BibitemShut {NoStop}%
\bibitem [{\citenamefont {Willingale}\ \emph {et~al.}(2010)\citenamefont
  {Willingale}, \citenamefont {Thomas}, \citenamefont {Nilson}, \citenamefont
  {Kaluza}, \citenamefont {Bandyopadhyay}, \citenamefont {Dangor},
  \citenamefont {Evans}, \citenamefont {Fernandes}, \citenamefont {Haines},
  \citenamefont {Kamperidis}, \citenamefont {Kingham}, \citenamefont {Minardi},
  \citenamefont {Notley}, \citenamefont {Ridgers}, \citenamefont {Rozmus},
  \citenamefont {Sherlock}, \citenamefont {Tatarakis}, \citenamefont {Wei},
  \citenamefont {Najmudin},\ and\ \citenamefont
  {Krushelnick}}]{willingale_fast_2010}%
  \BibitemOpen
  \bibfield  {author} {\bibinfo {author} {\bibfnamefont {L.}~\bibnamefont
  {Willingale}}, \bibinfo {author} {\bibfnamefont {A.~G.~R.}\ \bibnamefont
  {Thomas}}, \bibinfo {author} {\bibfnamefont {P.~M.}\ \bibnamefont {Nilson}},
  \bibinfo {author} {\bibfnamefont {M.~C.}\ \bibnamefont {Kaluza}}, \bibinfo
  {author} {\bibfnamefont {S.}~\bibnamefont {Bandyopadhyay}}, \bibinfo {author}
  {\bibfnamefont {A.~E.}\ \bibnamefont {Dangor}}, \bibinfo {author}
  {\bibfnamefont {R.~G.}\ \bibnamefont {Evans}}, \bibinfo {author}
  {\bibfnamefont {P.}~\bibnamefont {Fernandes}}, \bibinfo {author}
  {\bibfnamefont {M.~G.}\ \bibnamefont {Haines}}, \bibinfo {author}
  {\bibfnamefont {C.}~\bibnamefont {Kamperidis}}, \bibinfo {author}
  {\bibfnamefont {R.~J.}\ \bibnamefont {Kingham}}, \bibinfo {author}
  {\bibfnamefont {S.}~\bibnamefont {Minardi}}, \bibinfo {author} {\bibfnamefont
  {M.}~\bibnamefont {Notley}}, \bibinfo {author} {\bibfnamefont {C.~P.}\
  \bibnamefont {Ridgers}}, \bibinfo {author} {\bibfnamefont {W.}~\bibnamefont
  {Rozmus}}, \bibinfo {author} {\bibfnamefont {M.}~\bibnamefont {Sherlock}},
  \bibinfo {author} {\bibfnamefont {M.}~\bibnamefont {Tatarakis}}, \bibinfo
  {author} {\bibfnamefont {M.~S.}\ \bibnamefont {Wei}}, \bibinfo {author}
  {\bibfnamefont {Z.}~\bibnamefont {Najmudin}}, \ and\ \bibinfo {author}
  {\bibfnamefont {K.}~\bibnamefont {Krushelnick}},\ }\href {\doibase
  10.1103/PhysRevLett.105.095001} {\bibfield  {journal} {\bibinfo  {journal}
  {Phys. Rev. Lett.}\ }\textbf {\bibinfo {volume} {105}},\ \bibinfo {pages}
  {095001} (\bibinfo {year} {2010})}\BibitemShut {NoStop}%
\bibitem [{\citenamefont {Lancia}\ \emph {et~al.}(2014)\citenamefont {Lancia},
  \citenamefont {Albertazzi}, \citenamefont {Boniface}, \citenamefont
  {Grisollet}, \citenamefont {Riquier}, \citenamefont {Chaland}, \citenamefont
  {Le~Thanh}, \citenamefont {Mellor}, \citenamefont {Antici}, \citenamefont
  {Buffechoux}, \citenamefont {Chen}, \citenamefont {Doria}, \citenamefont
  {Nakatsutsumi}, \citenamefont {Peth}, \citenamefont {Swantusch},
  \citenamefont {Stardubtsev}, \citenamefont {Palumbo}, \citenamefont
  {Borghesi}, \citenamefont {Willi}, \citenamefont {P{\'e}pin},\ and\
  \citenamefont {Fuchs}}]{lancia_topology_2014}%
  \BibitemOpen
  \bibfield  {author} {\bibinfo {author} {\bibfnamefont {L.}~\bibnamefont
  {Lancia}}, \bibinfo {author} {\bibfnamefont {B.}~\bibnamefont {Albertazzi}},
  \bibinfo {author} {\bibfnamefont {C.}~\bibnamefont {Boniface}}, \bibinfo
  {author} {\bibfnamefont {A.}~\bibnamefont {Grisollet}}, \bibinfo {author}
  {\bibfnamefont {R.}~\bibnamefont {Riquier}}, \bibinfo {author} {\bibfnamefont
  {F.}~\bibnamefont {Chaland}}, \bibinfo {author} {\bibfnamefont {K.-C.}\
  \bibnamefont {Le~Thanh}}, \bibinfo {author} {\bibfnamefont {P.}~\bibnamefont
  {Mellor}}, \bibinfo {author} {\bibfnamefont {P.}~\bibnamefont {Antici}},
  \bibinfo {author} {\bibfnamefont {S.}~\bibnamefont {Buffechoux}}, \bibinfo
  {author} {\bibfnamefont {S.~N.}\ \bibnamefont {Chen}}, \bibinfo {author}
  {\bibfnamefont {D.}~\bibnamefont {Doria}}, \bibinfo {author} {\bibfnamefont
  {M.}~\bibnamefont {Nakatsutsumi}}, \bibinfo {author} {\bibfnamefont
  {C.}~\bibnamefont {Peth}}, \bibinfo {author} {\bibfnamefont {M.}~\bibnamefont
  {Swantusch}}, \bibinfo {author} {\bibfnamefont {M.}~\bibnamefont
  {Stardubtsev}}, \bibinfo {author} {\bibfnamefont {L.}~\bibnamefont
  {Palumbo}}, \bibinfo {author} {\bibfnamefont {M.}~\bibnamefont {Borghesi}},
  \bibinfo {author} {\bibfnamefont {O.}~\bibnamefont {Willi}}, \bibinfo
  {author} {\bibfnamefont {H.}~\bibnamefont {P{\'e}pin}}, \ and\ \bibinfo
  {author} {\bibfnamefont {J.}~\bibnamefont {Fuchs}},\ }\href {\doibase
  10.1103/PhysRevLett.113.235001} {\bibfield  {journal} {\bibinfo  {journal}
  {Phys. Rev. Lett.}\ }\textbf {\bibinfo {volume} {113}},\ \bibinfo {pages}
  {235001} (\bibinfo {year} {2014})}\BibitemShut {NoStop}%
\bibitem [{\citenamefont {Gao}\ \emph {et~al.}(2015)\citenamefont {Gao},
  \citenamefont {Nilson}, \citenamefont {Igumenshchev}, \citenamefont {Haines},
  \citenamefont {Froula}, \citenamefont {Betti},\ and\ \citenamefont
  {Meyerhofer}}]{gao_precision_2015}%
  \BibitemOpen
  \bibfield  {author} {\bibinfo {author} {\bibfnamefont {L.}~\bibnamefont
  {Gao}}, \bibinfo {author} {\bibfnamefont {P.~M.}\ \bibnamefont {Nilson}},
  \bibinfo {author} {\bibfnamefont {I.~V.}\ \bibnamefont {Igumenshchev}},
  \bibinfo {author} {\bibfnamefont {M.~G.}\ \bibnamefont {Haines}}, \bibinfo
  {author} {\bibfnamefont {D.~H.}\ \bibnamefont {Froula}}, \bibinfo {author}
  {\bibfnamefont {R.}~\bibnamefont {Betti}}, \ and\ \bibinfo {author}
  {\bibfnamefont {D.~D.}\ \bibnamefont {Meyerhofer}},\ }\href {\doibase
  10.1103/PhysRevLett.114.215003} {\bibfield  {journal} {\bibinfo  {journal}
  {Phys. Rev. Lett.}\ }\textbf {\bibinfo {volume} {114}},\ \bibinfo {pages}
  {215003} (\bibinfo {year} {2015})}\BibitemShut {NoStop}%
\bibitem [{\citenamefont {Bell}\ \emph {et~al.}(1981)\citenamefont {Bell},
  \citenamefont {Evans},\ and\ \citenamefont {Nicholas}}]{bell_elecron_1981}%
  \BibitemOpen
  \bibfield  {author} {\bibinfo {author} {\bibfnamefont {A.~R.}\ \bibnamefont
  {Bell}}, \bibinfo {author} {\bibfnamefont {R.~G.}\ \bibnamefont {Evans}}, \
  and\ \bibinfo {author} {\bibfnamefont {D.~J.}\ \bibnamefont {Nicholas}},\
  }\href {\doibase 10.1103/PhysRevLett.46.243} {\bibfield  {journal} {\bibinfo
  {journal} {Phys. Rev. Lett.}\ }\textbf {\bibinfo {volume} {46}},\ \bibinfo
  {pages} {243} (\bibinfo {year} {1981})}\BibitemShut {NoStop}%
\bibitem [{\citenamefont {Luciani}\ \emph {et~al.}(1985)\citenamefont
  {Luciani}, \citenamefont {Mora},\ and\ \citenamefont
  {Bendib}}]{luciani_magnetic_1985}%
  \BibitemOpen
  \bibfield  {author} {\bibinfo {author} {\bibfnamefont {J.~F.}\ \bibnamefont
  {Luciani}}, \bibinfo {author} {\bibfnamefont {P.}~\bibnamefont {Mora}}, \
  and\ \bibinfo {author} {\bibfnamefont {A.}~\bibnamefont {Bendib}},\ }\href
  {\doibase 10.1103/PhysRevLett.55.2421} {\bibfield  {journal} {\bibinfo
  {journal} {Phys. Rev. Lett.}\ }\textbf {\bibinfo {volume} {55}},\ \bibinfo
  {pages} {2421} (\bibinfo {year} {1985})}\BibitemShut {NoStop}%
\bibitem [{\citenamefont {Nilson}\ \emph {et~al.}(2006)\citenamefont {Nilson},
  \citenamefont {Willingale}, \citenamefont {Kaluza}, \citenamefont
  {Kamperidis}, \citenamefont {Minardi}, \citenamefont {Wei}, \citenamefont
  {Fernandes}, \citenamefont {Notley}, \citenamefont {Bandyopadhyay},
  \citenamefont {Sherlock}, \citenamefont {Kingham}, \citenamefont {Tatarakis},
  \citenamefont {Najmudin}, \citenamefont {Rozmus}, \citenamefont {Evans},
  \citenamefont {Haines}, \citenamefont {Dangor},\ and\ \citenamefont
  {Krushelnick}}]{nilson_magnetic_2006}%
  \BibitemOpen
  \bibfield  {author} {\bibinfo {author} {\bibfnamefont {P.~M.}\ \bibnamefont
  {Nilson}}, \bibinfo {author} {\bibfnamefont {L.}~\bibnamefont {Willingale}},
  \bibinfo {author} {\bibfnamefont {M.~C.}\ \bibnamefont {Kaluza}}, \bibinfo
  {author} {\bibfnamefont {C.}~\bibnamefont {Kamperidis}}, \bibinfo {author}
  {\bibfnamefont {S.}~\bibnamefont {Minardi}}, \bibinfo {author} {\bibfnamefont
  {M.~S.}\ \bibnamefont {Wei}}, \bibinfo {author} {\bibfnamefont
  {P.}~\bibnamefont {Fernandes}}, \bibinfo {author} {\bibfnamefont
  {M.}~\bibnamefont {Notley}}, \bibinfo {author} {\bibfnamefont
  {S.}~\bibnamefont {Bandyopadhyay}}, \bibinfo {author} {\bibfnamefont
  {M.}~\bibnamefont {Sherlock}}, \bibinfo {author} {\bibfnamefont {R.~J.}\
  \bibnamefont {Kingham}}, \bibinfo {author} {\bibfnamefont {M.}~\bibnamefont
  {Tatarakis}}, \bibinfo {author} {\bibfnamefont {Z.}~\bibnamefont {Najmudin}},
  \bibinfo {author} {\bibfnamefont {W.}~\bibnamefont {Rozmus}}, \bibinfo
  {author} {\bibfnamefont {R.~G.}\ \bibnamefont {Evans}}, \bibinfo {author}
  {\bibfnamefont {M.~G.}\ \bibnamefont {Haines}}, \bibinfo {author}
  {\bibfnamefont {A.~E.}\ \bibnamefont {Dangor}}, \ and\ \bibinfo {author}
  {\bibfnamefont {K.}~\bibnamefont {Krushelnick}},\ }\href {\doibase
  10.1103/PhysRevLett.97.255001} {\bibfield  {journal} {\bibinfo  {journal}
  {Phys. Rev. Lett.}\ }\textbf {\bibinfo {volume} {97}},\ \bibinfo {pages}
  {255001} (\bibinfo {year} {2006})}\BibitemShut {NoStop}%
\bibitem [{\citenamefont {Li}\ \emph {et~al.}(2007)\citenamefont {Li},
  \citenamefont {S{\'e}guin}, \citenamefont {Frenje}, \citenamefont {Rygg},
  \citenamefont {Petrasso}, \citenamefont {Town}, \citenamefont {Landen},
  \citenamefont {Knauer},\ and\ \citenamefont {Smalyuk}}]{li_observation_2007}%
  \BibitemOpen
  \bibfield  {author} {\bibinfo {author} {\bibfnamefont {C.~K.}\ \bibnamefont
  {Li}}, \bibinfo {author} {\bibfnamefont {F.~H.}\ \bibnamefont {S{\'e}guin}},
  \bibinfo {author} {\bibfnamefont {J.~A.}\ \bibnamefont {Frenje}}, \bibinfo
  {author} {\bibfnamefont {J.~R.}\ \bibnamefont {Rygg}}, \bibinfo {author}
  {\bibfnamefont {R.~D.}\ \bibnamefont {Petrasso}}, \bibinfo {author}
  {\bibfnamefont {R.~P.~J.}\ \bibnamefont {Town}}, \bibinfo {author}
  {\bibfnamefont {O.~L.}\ \bibnamefont {Landen}}, \bibinfo {author}
  {\bibfnamefont {J.~P.}\ \bibnamefont {Knauer}}, \ and\ \bibinfo {author}
  {\bibfnamefont {V.~A.}\ \bibnamefont {Smalyuk}},\ }\href {\doibase
  10.1103/PhysRevLett.99.055001} {\bibfield  {journal} {\bibinfo  {journal}
  {Phys. Rev. Lett.}\ }\textbf {\bibinfo {volume} {99}},\ \bibinfo {pages}
  {055001} (\bibinfo {year} {2007})}\BibitemShut {NoStop}%
\bibitem [{\citenamefont {Zhong}\ \emph {et~al.}(2010)\citenamefont {Zhong},
  \citenamefont {Li}, \citenamefont {Wang}, \citenamefont {Wang}, \citenamefont
  {Dong}, \citenamefont {Xiao}, \citenamefont {Wang}, \citenamefont {Liu},
  \citenamefont {Zhang}, \citenamefont {An}, \citenamefont {Wang},
  \citenamefont {Zhu}, \citenamefont {Gu}, \citenamefont {He}, \citenamefont
  {Zhao},\ and\ \citenamefont {Zhang}}]{zhong_modelling_2010}%
  \BibitemOpen
  \bibfield  {author} {\bibinfo {author} {\bibfnamefont {J.}~\bibnamefont
  {Zhong}}, \bibinfo {author} {\bibfnamefont {Y.}~\bibnamefont {Li}}, \bibinfo
  {author} {\bibfnamefont {X.}~\bibnamefont {Wang}}, \bibinfo {author}
  {\bibfnamefont {J.}~\bibnamefont {Wang}}, \bibinfo {author} {\bibfnamefont
  {Q.}~\bibnamefont {Dong}}, \bibinfo {author} {\bibfnamefont {C.}~\bibnamefont
  {Xiao}}, \bibinfo {author} {\bibfnamefont {S.}~\bibnamefont {Wang}}, \bibinfo
  {author} {\bibfnamefont {X.}~\bibnamefont {Liu}}, \bibinfo {author}
  {\bibfnamefont {L.}~\bibnamefont {Zhang}}, \bibinfo {author} {\bibfnamefont
  {L.}~\bibnamefont {An}}, \bibinfo {author} {\bibfnamefont {F.}~\bibnamefont
  {Wang}}, \bibinfo {author} {\bibfnamefont {J.}~\bibnamefont {Zhu}}, \bibinfo
  {author} {\bibfnamefont {Y.}~\bibnamefont {Gu}}, \bibinfo {author}
  {\bibfnamefont {X.}~\bibnamefont {He}}, \bibinfo {author} {\bibfnamefont
  {G.}~\bibnamefont {Zhao}}, \ and\ \bibinfo {author} {\bibfnamefont
  {J.}~\bibnamefont {Zhang}},\ }\href {\doibase 10.1038/nphys1790} {\bibfield
  {journal} {\bibinfo  {journal} {Nature Phys.}\ }\textbf {\bibinfo {volume}
  {6}},\ \bibinfo {pages} {984} (\bibinfo {year} {2010})}\BibitemShut {NoStop}%
\bibitem [{\citenamefont {Fox}\ \emph {et~al.}(2011)\citenamefont {Fox},
  \citenamefont {Bhattacharjee},\ and\ \citenamefont
  {Germaschewski}}]{fox_fast_2011}%
  \BibitemOpen
  \bibfield  {author} {\bibinfo {author} {\bibfnamefont {W.}~\bibnamefont
  {Fox}}, \bibinfo {author} {\bibfnamefont {A.}~\bibnamefont {Bhattacharjee}},
  \ and\ \bibinfo {author} {\bibfnamefont {K.}~\bibnamefont {Germaschewski}},\
  }\href {\doibase 10.1103/PhysRevLett.106.215003} {\bibfield  {journal}
  {\bibinfo  {journal} {Phys. Rev. Lett.}\ }\textbf {\bibinfo {volume} {106}},\
  \bibinfo {pages} {215003} (\bibinfo {year} {2011})}\BibitemShut {NoStop}%
\bibitem [{\citenamefont {Fox}\ \emph {et~al.}(2012)\citenamefont {Fox},
  \citenamefont {Bhattacharjee},\ and\ \citenamefont
  {Germaschewski}}]{fox_magnetic_2012}%
  \BibitemOpen
  \bibfield  {author} {\bibinfo {author} {\bibfnamefont {W.}~\bibnamefont
  {Fox}}, \bibinfo {author} {\bibfnamefont {A.}~\bibnamefont {Bhattacharjee}},
  \ and\ \bibinfo {author} {\bibfnamefont {K.}~\bibnamefont {Germaschewski}},\
  }\href {\doibase 10.1063/1.3694119} {\bibfield  {journal} {\bibinfo
  {journal} {Phys. Plasmas}\ }\textbf {\bibinfo {volume} {19}},\ \bibinfo
  {pages} {056309} (\bibinfo {year} {2012})}\BibitemShut {NoStop}%
\bibitem [{\citenamefont {Joglekar}\ \emph {et~al.}(2014)\citenamefont
  {Joglekar}, \citenamefont {Thomas}, \citenamefont {Fox},\ and\ \citenamefont
  {Bhattacharjee}}]{joglekar_magnetic_2014}%
  \BibitemOpen
  \bibfield  {author} {\bibinfo {author} {\bibfnamefont {A.~S.}\ \bibnamefont
  {Joglekar}}, \bibinfo {author} {\bibfnamefont {A.~G.~R.}\ \bibnamefont
  {Thomas}}, \bibinfo {author} {\bibfnamefont {W.}~\bibnamefont {Fox}}, \ and\
  \bibinfo {author} {\bibfnamefont {A.}~\bibnamefont {Bhattacharjee}},\ }\href
  {\doibase 10.1103/PhysRevLett.112.105004} {\bibfield  {journal} {\bibinfo
  {journal} {Phys. Rev. Lett.}\ }\textbf {\bibinfo {volume} {112}},\ \bibinfo
  {pages} {105004} (\bibinfo {year} {2014})}\BibitemShut {NoStop}%
\bibitem [{\citenamefont {Hesse}\ \emph {et~al.}(2001)\citenamefont {Hesse},
  \citenamefont {Birn},\ and\ \citenamefont
  {Kuznetsova}}]{hesse_collisionless_2001}%
  \BibitemOpen
  \bibfield  {author} {\bibinfo {author} {\bibfnamefont {M.}~\bibnamefont
  {Hesse}}, \bibinfo {author} {\bibfnamefont {J.}~\bibnamefont {Birn}}, \ and\
  \bibinfo {author} {\bibfnamefont {M.}~\bibnamefont {Kuznetsova}},\ }\href
  {\doibase 10.1029/1999JA001002} {\bibfield  {journal} {\bibinfo  {journal}
  {J. Geophys. Res.}\ }\textbf {\bibinfo {volume} {106}},\ \bibinfo {pages}
  {3721} (\bibinfo {year} {2001})}\BibitemShut {NoStop}%
\bibitem [{\citenamefont {Bessho}\ and\ \citenamefont
  {Bhattacharjee}(2005)}]{bessho_collisionless_2005}%
  \BibitemOpen
  \bibfield  {author} {\bibinfo {author} {\bibfnamefont {N.}~\bibnamefont
  {Bessho}}\ and\ \bibinfo {author} {\bibfnamefont {A.}~\bibnamefont
  {Bhattacharjee}},\ }\href {\doibase 10.1103/PhysRevLett.95.245001} {\bibfield
   {journal} {\bibinfo  {journal} {Phys. Rev. Lett.}\ }\textbf {\bibinfo
  {volume} {95}},\ \bibinfo {pages} {245001} (\bibinfo {year}
  {2005})}\BibitemShut {NoStop}%
\bibitem [{\citenamefont {Wang}\ \emph {et~al.}(2015)\citenamefont {Wang},
  \citenamefont {Hakim}, \citenamefont {Bhattacharjee},\ and\ \citenamefont
  {Germaschewski}}]{wang_comparison_2015}%
  \BibitemOpen
  \bibfield  {author} {\bibinfo {author} {\bibfnamefont {L.}~\bibnamefont
  {Wang}}, \bibinfo {author} {\bibfnamefont {A.~H.}\ \bibnamefont {Hakim}},
  \bibinfo {author} {\bibfnamefont {A.}~\bibnamefont {Bhattacharjee}}, \ and\
  \bibinfo {author} {\bibfnamefont {K.}~\bibnamefont {Germaschewski}},\ }\href
  {\doibase 10.1063/1.4906063} {\bibfield  {journal} {\bibinfo  {journal}
  {Phys. Plasmas}\ }\textbf {\bibinfo {volume} {22}},\ \bibinfo {pages}
  {012108} (\bibinfo {year} {2015})}\BibitemShut {NoStop}%
\bibitem [{\citenamefont {Catto}\ and\ \citenamefont
  {Simakov}(2004)}]{catto_drift_2004}%
  \BibitemOpen
  \bibfield  {author} {\bibinfo {author} {\bibfnamefont {P.~J.}\ \bibnamefont
  {Catto}}\ and\ \bibinfo {author} {\bibfnamefont {A.~N.}\ \bibnamefont
  {Simakov}},\ }\href {\doibase 10.1063/1.1632496} {\bibfield  {journal}
  {\bibinfo  {journal} {Physics of Plasmas (1994-present)}\ }\textbf {\bibinfo
  {volume} {11}},\ \bibinfo {pages} {90} (\bibinfo {year} {2004})}\BibitemShut
  {NoStop}%
\bibitem [{\citenamefont {Liu}\ \emph {et~al.}(2015)\citenamefont {Liu},
  \citenamefont {Fox},\ and\ \citenamefont {Bhattacharjee}}]{liu_heat_2015}%
  \BibitemOpen
  \bibfield  {author} {\bibinfo {author} {\bibfnamefont {C.}~\bibnamefont
  {Liu}}, \bibinfo {author} {\bibfnamefont {W.}~\bibnamefont {Fox}}, \ and\
  \bibinfo {author} {\bibfnamefont {A.}~\bibnamefont {Bhattacharjee}},\ }\href
  {\doibase 10.1063/1.4918941} {\bibfield  {journal} {\bibinfo  {journal}
  {Phys. Plasmas}\ }\textbf {\bibinfo {volume} {22}},\ \bibinfo {pages}
  {053302} (\bibinfo {year} {2015})}\BibitemShut {NoStop}%
\bibitem [{\citenamefont {Joglekar}\ \emph {et~al.}(2016)\citenamefont
  {Joglekar}, \citenamefont {Ridgers}, \citenamefont {Kingham},\ and\
  \citenamefont {Thomas}}]{joglekar_kinetic_2016}%
  \BibitemOpen
  \bibfield  {author} {\bibinfo {author} {\bibfnamefont {A.~S.}\ \bibnamefont
  {Joglekar}}, \bibinfo {author} {\bibfnamefont {C.~P.}\ \bibnamefont
  {Ridgers}}, \bibinfo {author} {\bibfnamefont {R.~J.}\ \bibnamefont
  {Kingham}}, \ and\ \bibinfo {author} {\bibfnamefont {A.~G.~R.}\ \bibnamefont
  {Thomas}},\ }\href {\doibase 10.1103/PhysRevE.93.043206} {\bibfield
  {journal} {\bibinfo  {journal} {Phys. Rev. E}\ }\textbf {\bibinfo {volume}
  {93}},\ \bibinfo {pages} {043206} (\bibinfo {year} {2016})}\BibitemShut
  {NoStop}%
\bibitem [{Note1()}]{Note1}%
  \BibitemOpen
  \bibinfo {note} {We will see later in the simulation results that inertial
  term is very small.}\BibitemShut {Stop}%
\bibitem [{\citenamefont {Braginskii}(1965)}]{braginskii_transport_1965}%
  \BibitemOpen
  \bibfield  {author} {\bibinfo {author} {\bibfnamefont {S.~I.}\ \bibnamefont
  {Braginskii}},\ }\href@noop {} {\bibfield  {journal} {\bibinfo  {journal}
  {Rev. Plasma Phys.}\ }\textbf {\bibinfo {volume} {1}},\ \bibinfo {pages}
  {205} (\bibinfo {year} {1965})}\BibitemShut {NoStop}%
\bibitem [{\citenamefont {Epperlein}\ and\ \citenamefont
  {Haines}(1986)}]{epperlein_plasma_1986}%
  \BibitemOpen
  \bibfield  {author} {\bibinfo {author} {\bibfnamefont {E.~M.}\ \bibnamefont
  {Epperlein}}\ and\ \bibinfo {author} {\bibfnamefont {M.~G.}\ \bibnamefont
  {Haines}},\ }\href {\doibase doi:10.1063/1.865901} {\bibfield  {journal}
  {\bibinfo  {journal} {Phys. Fluids}\ }\textbf {\bibinfo {volume} {29}},\
  \bibinfo {pages} {1029} (\bibinfo {year} {1986})}\BibitemShut {NoStop}%
\bibitem [{Note2()}]{Note2}%
  \BibitemOpen
  \bibinfo {note} {The result of $R_{3}$ indicates that for strongly magnetized
  case ($\Omega \tau \gg 1$) the layer width may be much smaller than $\lambda
  _{\protect \mathrm {mfp}}$. However, we found this is not true because the
  \protect \emph {additional} non-local transport effect, which will be
  discussed in Sec. \ref {sec:pic}, will introduce additional corrections to
  the amplitudes of the two terms.}\BibitemShut {Stop}%
\bibitem [{\citenamefont {Germaschewski}\ \emph {et~al.}(2016)\citenamefont
  {Germaschewski}, \citenamefont {Fox}, \citenamefont {Abbott}, \citenamefont
  {Ahmadi}, \citenamefont {Maynard}, \citenamefont {Wang}, \citenamefont
  {Ruhl},\ and\ \citenamefont {Bhattacharjee}}]{germaschewski_plasma_2016}%
  \BibitemOpen
  \bibfield  {author} {\bibinfo {author} {\bibfnamefont {K.}~\bibnamefont
  {Germaschewski}}, \bibinfo {author} {\bibfnamefont {W.}~\bibnamefont {Fox}},
  \bibinfo {author} {\bibfnamefont {S.}~\bibnamefont {Abbott}}, \bibinfo
  {author} {\bibfnamefont {N.}~\bibnamefont {Ahmadi}}, \bibinfo {author}
  {\bibfnamefont {K.}~\bibnamefont {Maynard}}, \bibinfo {author} {\bibfnamefont
  {L.}~\bibnamefont {Wang}}, \bibinfo {author} {\bibfnamefont {H.}~\bibnamefont
  {Ruhl}}, \ and\ \bibinfo {author} {\bibfnamefont {A.}~\bibnamefont
  {Bhattacharjee}},\ }\href {\doibase 10.1016/j.jcp.2016.05.013} {\bibfield
  {journal} {\bibinfo  {journal} {J. Comput. Phys.}\ }\textbf {\bibinfo
  {volume} {318}},\ \bibinfo {pages} {305} (\bibinfo {year}
  {2016})}\BibitemShut {NoStop}%
\bibitem [{Note3()}]{Note3}%
  \BibitemOpen
  \bibinfo {note} {Simulations with moving ions show nearly identical
  results.}\BibitemShut {Stop}%
\bibitem [{\citenamefont {Daughton}\ \emph {et~al.}(2009)\citenamefont
  {Daughton}, \citenamefont {Roytershteyn}, \citenamefont {Albright},
  \citenamefont {Karimabadi}, \citenamefont {Yin},\ and\ \citenamefont
  {Bowers}}]{daughton_influence_2009}%
  \BibitemOpen
  \bibfield  {author} {\bibinfo {author} {\bibfnamefont {W.}~\bibnamefont
  {Daughton}}, \bibinfo {author} {\bibfnamefont {V.}~\bibnamefont
  {Roytershteyn}}, \bibinfo {author} {\bibfnamefont {B.~J.}\ \bibnamefont
  {Albright}}, \bibinfo {author} {\bibfnamefont {H.}~\bibnamefont
  {Karimabadi}}, \bibinfo {author} {\bibfnamefont {L.}~\bibnamefont {Yin}}, \
  and\ \bibinfo {author} {\bibfnamefont {K.~J.}\ \bibnamefont {Bowers}},\
  }\href {\doibase doi:10.1063/1.3191718} {\bibfield  {journal} {\bibinfo
  {journal} {Phys. Plasmas}\ }\textbf {\bibinfo {volume} {16}},\ \bibinfo
  {pages} {072117} (\bibinfo {year} {2009})}\BibitemShut {NoStop}%
\bibitem [{\citenamefont {Epperlein}(1985)}]{epperlein_comparison_1985}%
  \BibitemOpen
  \bibfield  {author} {\bibinfo {author} {\bibfnamefont {E.~M.}\ \bibnamefont
  {Epperlein}},\ }\href {\doibase 10.1088/0741-3335/27/9/008} {\bibfield
  {journal} {\bibinfo  {journal} {Plasma Phys. Control. Fusion}\ }\textbf
  {\bibinfo {volume} {27}},\ \bibinfo {pages} {1027} (\bibinfo {year}
  {1985})}\BibitemShut {NoStop}%
\bibitem [{\citenamefont {Thomas}\ \emph {et~al.}(2009)\citenamefont {Thomas},
  \citenamefont {Kingham},\ and\ \citenamefont {Ridgers}}]{thomas_rapid_2009}%
  \BibitemOpen
  \bibfield  {author} {\bibinfo {author} {\bibfnamefont {A.~G.~R.}\
  \bibnamefont {Thomas}}, \bibinfo {author} {\bibfnamefont {R.~J.}\
  \bibnamefont {Kingham}}, \ and\ \bibinfo {author} {\bibfnamefont {C.~P.}\
  \bibnamefont {Ridgers}},\ }\href {\doibase 10.1088/1367-2630/11/3/033001}
  {\bibfield  {journal} {\bibinfo  {journal} {New J. Phys.}\ }\textbf {\bibinfo
  {volume} {11}},\ \bibinfo {pages} {033001} (\bibinfo {year}
  {2009})}\BibitemShut {NoStop}%
\bibitem [{Note4()}]{Note4}%
  \BibitemOpen
  \bibinfo {note} {The strong collisional-Weibel instability is partly due to
  the small-sized box in our PIC simulation due to the limitation of
  computation power. In a larger box the instability is expected to be much
  weaker thanks to the less focused heating region and smaller heat
  flux.}\BibitemShut {Stop}%
\bibitem [{\citenamefont {Dong}\ \emph {et~al.}(2012)\citenamefont {Dong},
  \citenamefont {Wang}, \citenamefont {Lu}, \citenamefont {Huang},
  \citenamefont {Yuan}, \citenamefont {Liu}, \citenamefont {Lin}, \citenamefont
  {Li}, \citenamefont {Wei}, \citenamefont {Zhong}, \citenamefont {Shi},
  \citenamefont {Jiang}, \citenamefont {Ding}, \citenamefont {Jiang},
  \citenamefont {Du}, \citenamefont {He}, \citenamefont {Yu}, \citenamefont
  {Liu}, \citenamefont {Wang}, \citenamefont {Tang}, \citenamefont {Zhu},
  \citenamefont {Zhao}, \citenamefont {Sheng},\ and\ \citenamefont
  {Zhang}}]{dong_plasmoid_2012}%
  \BibitemOpen
  \bibfield  {author} {\bibinfo {author} {\bibfnamefont {Q.-L.}\ \bibnamefont
  {Dong}}, \bibinfo {author} {\bibfnamefont {S.-J.}\ \bibnamefont {Wang}},
  \bibinfo {author} {\bibfnamefont {Q.-M.}\ \bibnamefont {Lu}}, \bibinfo
  {author} {\bibfnamefont {C.}~\bibnamefont {Huang}}, \bibinfo {author}
  {\bibfnamefont {D.-W.}\ \bibnamefont {Yuan}}, \bibinfo {author}
  {\bibfnamefont {X.}~\bibnamefont {Liu}}, \bibinfo {author} {\bibfnamefont
  {X.-X.}\ \bibnamefont {Lin}}, \bibinfo {author} {\bibfnamefont {Y.-T.}\
  \bibnamefont {Li}}, \bibinfo {author} {\bibfnamefont {H.-G.}\ \bibnamefont
  {Wei}}, \bibinfo {author} {\bibfnamefont {J.-Y.}\ \bibnamefont {Zhong}},
  \bibinfo {author} {\bibfnamefont {J.-R.}\ \bibnamefont {Shi}}, \bibinfo
  {author} {\bibfnamefont {S.-E.}\ \bibnamefont {Jiang}}, \bibinfo {author}
  {\bibfnamefont {Y.-K.}\ \bibnamefont {Ding}}, \bibinfo {author}
  {\bibfnamefont {B.-B.}\ \bibnamefont {Jiang}}, \bibinfo {author}
  {\bibfnamefont {K.}~\bibnamefont {Du}}, \bibinfo {author} {\bibfnamefont
  {X.-T.}\ \bibnamefont {He}}, \bibinfo {author} {\bibfnamefont {M.~Y.}\
  \bibnamefont {Yu}}, \bibinfo {author} {\bibfnamefont {C.~S.}\ \bibnamefont
  {Liu}}, \bibinfo {author} {\bibfnamefont {S.}~\bibnamefont {Wang}}, \bibinfo
  {author} {\bibfnamefont {Y.-J.}\ \bibnamefont {Tang}}, \bibinfo {author}
  {\bibfnamefont {J.-Q.}\ \bibnamefont {Zhu}}, \bibinfo {author} {\bibfnamefont
  {G.}~\bibnamefont {Zhao}}, \bibinfo {author} {\bibfnamefont {Z.-M.}\
  \bibnamefont {Sheng}}, \ and\ \bibinfo {author} {\bibfnamefont
  {J.}~\bibnamefont {Zhang}},\ }\href {\doibase 10.1103/PhysRevLett.108.215001}
  {\bibfield  {journal} {\bibinfo  {journal} {Phys. Rev. Lett.}\ }\textbf
  {\bibinfo {volume} {108}},\ \bibinfo {pages} {215001} (\bibinfo {year}
  {2012})}\BibitemShut {NoStop}%
\bibitem [{\citenamefont {Zhong}\ \emph {et~al.}(2016)\citenamefont {Zhong},
  \citenamefont {Lin}, \citenamefont {Li}, \citenamefont {Wang}, \citenamefont
  {Li}, \citenamefont {Zhang}, \citenamefont {Yuan}, \citenamefont {Ping},
  \citenamefont {Wei}, \citenamefont {Wang}, \citenamefont {Su}, \citenamefont
  {Li}, \citenamefont {Han}, \citenamefont {Liao}, \citenamefont {Yin},
  \citenamefont {Fang}, \citenamefont {Yuan}, \citenamefont {Wang},
  \citenamefont {Sun}, \citenamefont {Liang}, \citenamefont {Wang},
  \citenamefont {Ding}, \citenamefont {He}, \citenamefont {Zhu}, \citenamefont
  {Sheng}, \citenamefont {Li}, \citenamefont {Zhao},\ and\ \citenamefont
  {Zhang}}]{zhong_relativistic_2016}%
  \BibitemOpen
  \bibfield  {author} {\bibinfo {author} {\bibfnamefont {J.~Y.}\ \bibnamefont
  {Zhong}}, \bibinfo {author} {\bibfnamefont {J.}~\bibnamefont {Lin}}, \bibinfo
  {author} {\bibfnamefont {Y.~T.}\ \bibnamefont {Li}}, \bibinfo {author}
  {\bibfnamefont {X.}~\bibnamefont {Wang}}, \bibinfo {author} {\bibfnamefont
  {Y.}~\bibnamefont {Li}}, \bibinfo {author} {\bibfnamefont {K.}~\bibnamefont
  {Zhang}}, \bibinfo {author} {\bibfnamefont {D.~W.}\ \bibnamefont {Yuan}},
  \bibinfo {author} {\bibfnamefont {Y.~L.}\ \bibnamefont {Ping}}, \bibinfo
  {author} {\bibfnamefont {H.~G.}\ \bibnamefont {Wei}}, \bibinfo {author}
  {\bibfnamefont {J.~Q.}\ \bibnamefont {Wang}}, \bibinfo {author}
  {\bibfnamefont {L.~N.}\ \bibnamefont {Su}}, \bibinfo {author} {\bibfnamefont
  {F.}~\bibnamefont {Li}}, \bibinfo {author} {\bibfnamefont {B.}~\bibnamefont
  {Han}}, \bibinfo {author} {\bibfnamefont {G.~Q.}\ \bibnamefont {Liao}},
  \bibinfo {author} {\bibfnamefont {C.~L.}\ \bibnamefont {Yin}}, \bibinfo
  {author} {\bibfnamefont {Y.}~\bibnamefont {Fang}}, \bibinfo {author}
  {\bibfnamefont {X.}~\bibnamefont {Yuan}}, \bibinfo {author} {\bibfnamefont
  {C.}~\bibnamefont {Wang}}, \bibinfo {author} {\bibfnamefont {J.~R.}\
  \bibnamefont {Sun}}, \bibinfo {author} {\bibfnamefont {G.~Y.}\ \bibnamefont
  {Liang}}, \bibinfo {author} {\bibfnamefont {F.~L.}\ \bibnamefont {Wang}},
  \bibinfo {author} {\bibfnamefont {Y.~K.}\ \bibnamefont {Ding}}, \bibinfo
  {author} {\bibfnamefont {X.~T.}\ \bibnamefont {He}}, \bibinfo {author}
  {\bibfnamefont {J.~Q.}\ \bibnamefont {Zhu}}, \bibinfo {author} {\bibfnamefont
  {Z.~M.}\ \bibnamefont {Sheng}}, \bibinfo {author} {\bibfnamefont
  {G.}~\bibnamefont {Li}}, \bibinfo {author} {\bibfnamefont {G.}~\bibnamefont
  {Zhao}}, \ and\ \bibinfo {author} {\bibfnamefont {J.}~\bibnamefont {Zhang}},\
  }\href {\doibase 10.3847/0067-0049/225/2/30} {\bibfield  {journal} {\bibinfo
  {journal} {ApJS}\ }\textbf {\bibinfo {volume} {225}},\ \bibinfo {pages} {30}
  (\bibinfo {year} {2016})}\BibitemShut {NoStop}%
\bibitem [{\citenamefont {Dreicer}(1959)}]{dreicer_electron_1959}%
  \BibitemOpen
  \bibfield  {author} {\bibinfo {author} {\bibfnamefont {H.}~\bibnamefont
  {Dreicer}},\ }\href {\doibase 10.1103/PhysRev.115.238} {\bibfield  {journal}
  {\bibinfo  {journal} {Phys. Rev.}\ }\textbf {\bibinfo {volume} {115}},\
  \bibinfo {pages} {238} (\bibinfo {year} {1959})}\BibitemShut {NoStop}%
\bibitem [{\citenamefont {Luciani}\ \emph {et~al.}(1983)\citenamefont
  {Luciani}, \citenamefont {Mora},\ and\ \citenamefont
  {Virmont}}]{luciani_nonlocal_1983}%
  \BibitemOpen
  \bibfield  {author} {\bibinfo {author} {\bibfnamefont {J.~F.}\ \bibnamefont
  {Luciani}}, \bibinfo {author} {\bibfnamefont {P.}~\bibnamefont {Mora}}, \
  and\ \bibinfo {author} {\bibfnamefont {J.}~\bibnamefont {Virmont}},\ }\href
  {\doibase 10.1103/PhysRevLett.51.1664} {\bibfield  {journal} {\bibinfo
  {journal} {Phys. Rev. Lett.}\ }\textbf {\bibinfo {volume} {51}},\ \bibinfo
  {pages} {1664} (\bibinfo {year} {1983})}\BibitemShut {NoStop}%
\bibitem [{\citenamefont {Hammett}\ and\ \citenamefont
  {Perkins}(1990)}]{hammett_fluid_1990}%
  \BibitemOpen
  \bibfield  {author} {\bibinfo {author} {\bibfnamefont {G.~W.}\ \bibnamefont
  {Hammett}}\ and\ \bibinfo {author} {\bibfnamefont {F.~W.}\ \bibnamefont
  {Perkins}},\ }\href {\doibase 10.1103/PhysRevLett.64.3019} {\bibfield
  {journal} {\bibinfo  {journal} {Phys. Rev. Lett.}\ }\textbf {\bibinfo
  {volume} {64}},\ \bibinfo {pages} {3019} (\bibinfo {year}
  {1990})}\BibitemShut {NoStop}%
\bibitem [{\citenamefont {Snyder}\ \emph {et~al.}(1997)\citenamefont {Snyder},
  \citenamefont {Hammett},\ and\ \citenamefont {Dorland}}]{snyder_landau_1997}%
  \BibitemOpen
  \bibfield  {author} {\bibinfo {author} {\bibfnamefont {P.~B.}\ \bibnamefont
  {Snyder}}, \bibinfo {author} {\bibfnamefont {G.~W.}\ \bibnamefont {Hammett}},
  \ and\ \bibinfo {author} {\bibfnamefont {W.}~\bibnamefont {Dorland}},\ }\href
  {\doibase 10.1063/1.872517} {\bibfield  {journal} {\bibinfo  {journal} {Phys.
  Plasmas}\ }\textbf {\bibinfo {volume} {4}},\ \bibinfo {pages} {3974}
  (\bibinfo {year} {1997})}\BibitemShut {NoStop}%
\bibitem [{\citenamefont {Thomas}\ \emph {et~al.}(2012)\citenamefont {Thomas},
  \citenamefont {Tzoufras}, \citenamefont {Robinson}, \citenamefont {Kingham},
  \citenamefont {Ridgers}, \citenamefont {Sherlock},\ and\ \citenamefont
  {Bell}}]{thomas_review_2012}%
  \BibitemOpen
  \bibfield  {author} {\bibinfo {author} {\bibfnamefont {A.~G.~R.}\
  \bibnamefont {Thomas}}, \bibinfo {author} {\bibfnamefont {M.}~\bibnamefont
  {Tzoufras}}, \bibinfo {author} {\bibfnamefont {A.~P.~L.}\ \bibnamefont
  {Robinson}}, \bibinfo {author} {\bibfnamefont {R.~J.}\ \bibnamefont
  {Kingham}}, \bibinfo {author} {\bibfnamefont {C.~P.}\ \bibnamefont
  {Ridgers}}, \bibinfo {author} {\bibfnamefont {M.}~\bibnamefont {Sherlock}}, \
  and\ \bibinfo {author} {\bibfnamefont {A.~R.}\ \bibnamefont {Bell}},\ }\href
  {\doibase 10.1016/j.jcp.2011.09.028} {\bibfield  {journal} {\bibinfo
  {journal} {J. Comput. Phys.}\ }\textbf {\bibinfo {volume} {231}},\ \bibinfo
  {pages} {1051} (\bibinfo {year} {2012})}\BibitemShut {NoStop}%
\bibitem [{\citenamefont {Shkarofsky}\ \emph {et~al.}(1966)\citenamefont
  {Shkarofsky}, \citenamefont {Johnston},\ and\ \citenamefont
  {Bachynski}}]{shkarofsky_particle_1966}%
  \BibitemOpen
  \bibfield  {author} {\bibinfo {author} {\bibfnamefont {I.~P.}\ \bibnamefont
  {Shkarofsky}}, \bibinfo {author} {\bibfnamefont {T.~W.}\ \bibnamefont
  {Johnston}}, \ and\ \bibinfo {author} {\bibfnamefont {M.~P.}\ \bibnamefont
  {Bachynski}},\ }\href@noop {} {\emph {\bibinfo {title} {The {Particle}
  {Kinetics} of {Plasmas}}}}\ (\bibinfo  {publisher} {Addison-Wesley},\
  \bibinfo {year} {1966})\BibitemShut {NoStop}%
\end{thebibliography}%

\appendix
\section{Derivations of the generalized Ohm's law}
\label{sec:appendix}

In this section, we show the calculation of the generalized Ohm's law from the kinetic equation. We follow standard technique in the transport theory, by doing an expansion of the distribution function $f$, and take moments of the  equation for $\mathbf{f}_{1}$. The kinetic equation with Landau collision operator, including both the test-particle and field-particle parts, is a complicated integrodifferental equation. To solve this equation,  we use a finite difference method to solve the kinetic equation numerically, in order to overcome the inaccuracies in the polynomial expansion method\cite{braginskii_transport_1965}. The method is similar to that applied in \cite{epperlein_plasma_1986}, but we include the contribution from the higher order term of the distribution function to take into account the first order nonlocal effect.

We now proceed with the calculation. The electrons distribution $f$ can be expanded in a Cartesian form,
\begin{equation}
f=f_{0}+\mathbf{f}_{1}\cdot \mathbf{v}/v+\underline{\underline{\mathsf{f}_{2}}}:\mathbf{v}\mathbf{v}/v^{2}.
\end{equation}
The kinetic equation for $\mathbf{f_1}$ can be written as follows\cite{shkarofsky_particle_1966},
\begin{align}
\label{chapman-enskog}
\frac{d\mathbf{f}_1}{dt}+v\nabla f_0-\frac{\mathbf{E}e}{m_{e}}\frac{\partial f_0}{\partial v}+\mathbf{\Omega}\times \mathbf{f}_1+\frac{2}{5}v\nabla \cdot \mathsf{f}_2\nonumber\\
-\frac{2}{5v^3}\frac{\partial}{\partial v}(v^3 \frac{\mathbf{E}e}{m_{e}}\cdot \mathsf{f}_2)=C_{1}[\mathbf{f}_1],
\end{align}
where $\mathbf{\Omega}$ is the electron cyclotron frequency with the same direction of $\mathbf{B}$. $C_{1}$ is the collision operator for $f_{1}$. For simplicity, we focus on electron kinetics and assume that ions are stationary. Note that the last two terms on the left-hand-side (LHS) describe the contribution of $\mathsf{f}_{2}$ to the evolution of $\mathbf{f}_{1}$, which are ignored in \cite{epperlein_plasma_1986}.

The collision operator $C_{1}$ can be derived from the Landau collision operator $C_{L}$ as $C_{1}=(1/4\pi)\int d\mathbf{v} \mathbf{v} C_{L}$\cite{shkarofsky_particle_1966,epperlein_plasma_1986}. $C_{1}$ has the same form the the three component of $\mathbf{f}_{1}$. Including both the electron-electron and the electron-ion collisions, the collision operator can be expressed as,
\begin{equation}
  C_{1}=C_{ee}+C_{ei},
\end{equation}
\begin{multline}
C_{ee}=n^{-1}\nu_{ee}\left[\frac{v^{2}}{3}(I_{2}^{0}+J_{-1}^{0})\frac{d^{2}f_{1}}{dv^{2}}\right.\\
+\frac{v}{3}(-I_{2}^{0}+2J_{-1}^{0}+3I_{0}^{0})\frac{df_{1}}{dv}+\frac{1}{3}(I_{2}^{0}-2J_{-1}^{0}-3I_{0}^{0})f_{1}\\
+8\pi v^{3}f_{1}f_{0}+\frac{v^{2}}{5}\left(I_{3}^{1}+J_{-2}^{1})\right)\frac{d^{2}f_{0}}{dv^{2}}\\
\left.+\frac{v}{15}(-3I_{3}^{1}+2J_{-2}^{1}+5I_{1}^{1})\frac{df_{0}}{dv}\right],
\end{multline}
\begin{equation}
\label{cei}
C_{ei}=-\nu_{ei}f_{1},
\end{equation}
where $\nu_{ee}=\left[4\pi n_e(e^{2}/m_e)^{2}\ln \Lambda\right]/v^3$, $\nu_{ei}=\left[4\pi n_{i}(Ze^{2}/m_e)^{2}\ln \Lambda\right]/v^3$, and
\begin{equation}
I_{j}^{i}=4\pi v^{-j}\int_{0}^{v}f_{i}v^{2+j}dv,\quad J_{j}^{i}=4\pi v^{-j}\int_{v}^{\infty}f_{i}v^{j+2}dv.
\end{equation}

For a given $f_{0}$ and $\mathsf{f}_{2}$, the distribution function $\mathbf{f}_{1}$ can then be solved from Eq. (\ref{chapman-enskog}) using the finite difference method. The current and the generized Ohm's law can be obtained by taking moment of the resultant $\mathbf{f}_{1}$
\begin{equation}
\mathbf{j}=\frac{4\pi}{3}\int \mathbf{f}_{1}v^3dv.
\end{equation}

The value of $f_{0}$ is chosen as a Maxwellian distribution,
\begin{equation*}
f_0(W)=n_{e}\left(\frac{m_{e}}{2\pi T_{e}}\right)^{3/2}\exp(-W),\qquad W=\frac{m_{e}v^{2}}{2T_{e}},
\end{equation*}
where $n_{e}$ and $T_{e}$ can be inhomogeneous in space.

For $\mathsf{f}_{2}$, we note that the anisotropic pressure tensor is associated with the moment of it,
\begin{equation}
  \label{pi-tensor}
  {\mathsf{\Pi}}={\mathsf{P}}-p{\mathsf{I}}_2=\frac{8\pi m_{e}}{15}\int \mathsf{f}_2 v^4 dv.
\end{equation}
If we expand $\mathsf{f}_{2}$ using the generalized Laguerre polynomials like \cite{braginskii_transport_1965},
\begin{equation}
\mathsf{f}_{2}=v^{2}f_0(W)\sum_{r=0}^{\infty} \mathsf{R}_{r} L^{5/2}_r(W),
\end{equation}
and using the orthogonality relations of $L^{5/2}_{r}$, we find that the coefficient $\mathsf{R}_{0}$ corresponds to the anisotropic pressure tensor,
\begin{equation}
\mathsf{\Pi}=\frac{2n_{e}T_{e}^{2}}{m_{e}}\mathsf{R}_{0}.
\end{equation}

Vice versa, to examine how the pressure tensor affects  the evolution of $\mathbf{f}_{1}$, we can choose $\mathsf{f}_{2}$ which only contains the $L_{0}^{5/2}$ component. In this way we can calculate the transport coefficients associated with $\mathsf{\Pi}$ in the generalized Ohm's law.

The new Ohm's law derived from this procedure can be expressed as
\begin{align}
\label{ohm2}
\mathbf{E}=\frac{\underline{\underline\alpha}\cdot\mathbf{j}}{n_{e}^{2}e^{2}}+\frac{\mathbf{j}\times \mathbf{B}}{n_{e}ec}&-\frac{\nabla \cdot \left(p_{e}\mathsf{I}+\mathsf{\Pi}\right)}{n_{e} e}\nonumber\\
-\frac{\underline{\underline\beta}\cdot\nabla \cdot \left(T_{e}\mathsf{I}+\frac{2}{5}\mathsf{\Pi}/n_{e}\right)}{e}-\frac{\mathsf{\Pi}}{n_{e}e}&\cdot\underline{\underline{\gamma}}\cdot\frac{\nabla T_{e}}{T_{e}}-\frac{2\underline{\underline\beta}}{5}\cdot\mathbf{E}\cdot\frac{\mathsf{\Pi}}{p_{e}}.
\end{align}
Note that in addition to the terms in the Ohm's law in \cite{epperlein_plasma_1986}, our new generalized Ohm's law includes four addtional terms related to $\mathsf{\Pi}$. The first one shows as a divergence of momentum flow, which also appears in Eq. (\ref{ohm1}). The second one also depends on $\nabla\cdot\mathsf{\Pi}$, but actually comes from the friction force $\mathbf{R}_{ei}$ in Eq. (\ref{ohm1}). The transport coefficient of this term happens to be the same as the thermoelectric term $\beta\cdot\nabla T_{e}$ with an additional factor $2/5$. The last two terms on the right-hand-side are  ``cross-terms'' which depends on the inner product of $\mathsf{\Pi}$ and other transport forces ($\nabla T_{e}$, $\mathbf{E}$), where $\underline{\underline\gamma}$ is another transport coefficient depending on $\Omega\tau$ and $Z$. However, in this paper we ignore the effects of the cross terms, given that the gradient scale length of $\mathsf{\Pi}$ ($\mathsf{\Pi}/\nabla \mathsf{\Pi}$) is typically much shorter than that of pressure or temperature, and $\mathsf{\Pi}\ll p_{e}$, so the cross terms are subdominant compared to other terms in GOL.

\end{document}